\documentclass[fleqn,usenatbib]{mnras}

\usepackage{graphicx}	
\usepackage{amsmath}	

\usepackage{amssymb}	
\usepackage{multicol}       
\usepackage{bm}		
\usepackage{pdflscape}	
\usepackage[T1]{fontenc}
\usepackage{ae,aecompl}
\usepackage{newtxtext,newtxmath}
\usepackage{lineno} 
\usepackage{xurl}

\title[Neural Network Emulators: Forecasting Growth-Geometry split]{Attention-based Neural Network Emulators for Multi-Probe Data Vectors Part I: Forecasting the Growth-Geometry split}

\author[Kunhao Zhong et al.]{
Kunhao Zhong,$^{1, 2}$\thanks{kunhao.zhong@stonybrook.edu},
Evan Saraivanov$^{1}$,
James Caputi$^{1}$,
Vivian Miranda$^{3}$,
Supranta S. Boruah$^{4}$,
Tim Eifler$^{4}$,
\newauthor
Elisabeth Krause$^{4,5}$
\\
$^{1}$Department of Physics and Astronomy, Stony Brook University, Stony Brook, NY 11794, USA \\
$^{2}$Department of Physics and Astronomy,
University of Pennsylvania, Philadelphia, PA 19104, USA \\
$^{3}$C. N. Yang Institute for Theoretical Physics, 
Stony Brook University, Stony Brook, NY 11794, USA\\
$^{4}$Department of Astronomy and Steward Observatory, University of Arizona, 933 N Cherry Ave, Tucson, AZ 85719, USA\\
$^{5}$Department of Physics, University of Arizona,  1118 E Fourth Str, Tucson, AZ, 85721-0065, USA \\
}

\date{Accepted XXX. Received YYY; in original form ZZZ}

\pubyear{2024}

\begin{document}
\label{firstpage}
\pagerange{\pageref{firstpage}--\pageref{lastpage}}
\maketitle

\begin{abstract}
{
We present a new class of machine-learning emulators that accurately model the cosmic shear, galaxy-galaxy lensing, and galaxy clustering real space correlation functions in the context of Rubin Observatory year one simulated data. To illustrate its capabilities in forecasting models beyond the standard $\Lambda$CDM, we forecast how well LSST Year 1 data will be able to probe the consistency between geometry $\Omega^{\rm geo}_\mathrm{m}$ and growth $\Omega^{\rm growth}_\mathrm{m}$ dark matter densities in the so-called split $\Lambda$CDM parameterization. When trained with a few million samples, our emulator shows uniform accuracy across a wide range in an 18-dimensional parameter space. We provide a detailed comparison of three neural network designs, illustrating the importance of adopting state-of-the-art Transformer blocks. Our study also details their performance when computing Bayesian evidence for cosmic shear on three fiducial cosmologies. The transformers-based emulator is always accurate within \textsc{PolyChord}'s precision. As an application, we use our emulator to study the degeneracies between dark energy models and growth geometry split parameterizations. We find that the growth-geometry split remains to be a meaningful test of the smooth dark energy assumption.
}
\end{abstract}

\maketitle

\section{Introduction}

Markov Chain Monte Carlo (MCMC) is a common and effective technique for inferring cosmological parameters, as it enables the exploration of high-dimensional likelihoods from a wide range of data sets. These include space and ground-based measurements of the Cosmic Microwave Background (CMB)~\citep{Planck:2018nkj,ACT:2020gnv,SPT-3G:2021eoc}, type Ia supernova~\citep{Pan-STARRS1:2017jku,2022ApJ...938..113S}, Baryon Acoustic Oscillations (BAO)~\citep{Ross:2014qpa,BOSS:2016wmc,Raichoor:2020vio,eBOSS:2020yzd,2022MNRAS.511.5492Z}, and lensing and clustering of optical galaxies~\citep{HSC:2018mrq,Hamana:2019etx,LSST:2019wqx,KiDS:2020suj,Heymans:2020gsg,DES:2021wwk}.

Forthcoming stage-IV surveys will observe the lensing and clustering of optical galaxies with unparalleled precision. These experiments include the Dark Energy Spectroscopic Instrument~\citep{2023arXiv230606307D}, the Large Synoptic Survey Telescope (LSST)~\citep{LSST:2008ijt}, the Euclid mission~\citep{laureijs2011euclid}, and the Roman Space Telescope~\citep{Dore:2019pld}. They all aim to provide a more ambitious investigation into the nature of dark energy and dark matter, with a corresponding increase in the model complexity of nuisance physics. For example, new analyses will consider parameterizations for galaxy biases and intrinsic alignments that are better-motivated from an effective field theory perspective~\citep{Kokron:2021xgh,Mergulhao:2021kip,Bakx:2023mld,Chen:2023yyb,Nicola:2023hsd,Rubira:2023vzw}. The dark sector may simultaneously include new physics in the early~\citep{CarrilloGonzalez:2020oac,Benevento:2022cql,Hill:2021yec,Buen-Abad:2023uva,Berryman:2022hds,FrancoAbellan:2023gec,McDonough:2023qcu,Niedermann:2023ssr,Poulin:2023lkg,Ramadan:2023ivw}, intermediate~\citep{Simon:2022ftd,Holm:2022eqq,Alvi:2022aam,Anchordoqui:2022gmw,Ashoorioon:2023jwf,Bucko:2023eix,Holm:2022kkd,Nygaard:2023gel,Rubira:2022xhb}, and late Universes~\citep{Gleyzes:2015pma,Zumalacarregui:2016pph,Nesseris:2017vor,Peirone:2017ywi,Espejo:2018hxa,Sakr:2021ylx,Amon:2022azi,Berghaus:2023ypi,Lin:2023uux,Wang:2023tjj,Zhong:2023how}, as simple alternatives to $\Lambda$CDM is shown to be ineffective in explaining current tensions~\citep{Schoneberg:2021qvd}.

Future investigations into the dark sector and nuisance physics are expected to have higher dimensionality of the parameter space and larger execution time per model. For example, CMB constraints from the damping tail of the temperature power spectrum at multipoles $\ell \gtrsim 3000$ will soon require additional parameters related to baryonic physics~\citep{McCarthy:2021lfp}. The stiff increase in computational costs associated with running MCMC simulations is inevitable as the precision of theory predictions must align with the reduced uncertainties of new experiments.

Indeed, the current computational costs associated with evaluating some statistical metrics that assess tension and goodness-of-fit are already slowing the research progress. Currently, MCMCs aimed at constraining early dark energy scenarios, often employing a combination of CMB data measured by Planck and ACT experiments~\citep{Hill:2021yec}, can take weeks to satisfy the stringent $R - 1 \approx 0.005$ Gelman-Rubin convergence threshold~\citep{10.1214/ss/1177011136}. This criterion is necessary for robustly assessing goodness-of-fit~\citep{Raveri:2018wln}. Moreover, the marginalization of Early Dark Energy predictions over general late-time dark energy equations of state further increases the computational demands of MCMCs, with convergence being achieved only after months of continuous runs~\citep{Reboucas:2023rjm}. As the standard $\Lambda$CDM model undergoes simultaneous generalizations in the early and late times to potentially solve multiple existing tensions (or to stress test the standard model), these demanding analyses are expected to become more common.

Forecasts of the upcoming Simons Observatory, CMB-S4, and CMB-HD experiments have shown the need for enhancements in the accuracy settings of CLASS~\citep{CLASS-code} and CAMB~\citep{Lewis:2002ah,Howlett_2012} Boltzmann codes. These new settings
considerably increase the run time per model, even when predictions are restricted to the $\Lambda$CDM model~\citep{McCarthy:2021lfp,MacInnis:2023vif}. Exploring the viability of new theories with Bayesian tools may soon become impractical even when abundant computational resources are available, especially considering that interpreting tension metrics requires calibration with simulated data~\citep{Miranda:2020lpk}. 

A promising strategy to address these challenges lies in the development of accurate machine learning-based emulators for the theoretical data vectors; they have proven beneficial in a myriad of forecasts~\citep{2020MNRAS.491.2655M, 2022MNRAS.511.1771S, 2022JCAP...11..035G,Boruah:2022uac, 2023JCAP...01..016T}. The training and validation of emulators involve calculating tens of thousands, sometimes millions, of data vectors.  However, training is cost-effective compared to MCMCs since the evaluations can be trivially parallelized. There is also no need for the emulator to be able to extrapolate its predictions to regions in parameter space outside the boundary of training, which simplifies the underlying machine learning architecture and reduces the training time. 

Emulators based on the simpler multilayer perceptron machine learning design can accurately predict CMB and weak lensing data vectors in Fourier space over large volumes in parameter space~\citep{2022MNRAS.511.1771S}. In recent studies, \citet{Boruah:2022uac} and~\citet{2023JCAP...01..016T} successfully created emulators that compute the real-space cosmic shear, galaxy-galaxy lensing, and galaxy clustering two-point correlation functions of current- and next-generation of optical surveys based on widely adopted deep learning designs. In both cases, the complexity of data vectors in real space restricted the volume of parameter space in which these emulators could provide meaningful results. 

For instance,~\cite{Boruah:2022uac} trained the emulator using an adaptive method that is more or less equivalent to a temperate MCMC with a temperature of a few around an arbitrary cosmology (temperature is a scalar that broadens the covariance matrix). The resulting emulator fails to provide accurate results when the MCMC best-fit cosmology shifts more than a few sigmas away from the fiducial model assumed on training. This failing can happen whenever forecasting is performed around a different cosmology or in analysis with actual data if the high likelihood region is far from the fiducial training model. As long as their emulator is constantly retrained, and their training method is significantly cheaper than what we propose in this work, this limitation is not a significant problem, as both \citet{Boruah:2022uac} and~\citet{2023JCAP...01..016T} had the confined aim to create methods for approximate computing $68\%$ and $95\%$ confidence regions in MCMCs. 

Nevertheless, there are instances where it is advantageous to have emulators with a large volume of applicability in parameter space. When dealing with actual data, the combination of likelihoods from various experiments can shift parameters 
noticeably; this requires constant emulator retraining whenever the data combination changes (and there are a lot of possible combinations). Another case involves the computation of the Bayesian evidence, which is more sensitive to the modeling accuracy on the tails of the probability distribution. Various samplers are available to compute Bayesian evidence, including \textsc{MultiNest}~\citep{2008MNRAS.384..449F}, \textsc{PolyChord}~\citep{2015MNRAS.450L..61H}, \textsc{Dynesty}~\citep{2020MNRAS.493.3132S}, \textsc{PocoMC}~\citep{karamanis2022accelerating}, and \textsc{Nautilus}~\citep{Lange:2023ydq}. Each of them employs a different algorithm that contains hyperparameters that must be tuned to the target posterior distribution, necessitating calibration with synthetic data vectors that can be extremely expensive~\citep{Miranda:2020lpk,2023MNRAS.521.1184L}.

In this paper we study the growth geometry split, which is a phenomenological test that duplicates two parameters in wCDM cosmologies, the cold dark matter density $\Omega_\mathrm{m}$ and the dark energy equation of state $w$~\citep{Wang:2007fsa,2015PhRvD..91f3009R,DES:2020iqt,Ruiz-Zapatero:2021rzl,Nguyen:2023fip,Zhong:2023how}.  The additional parameters disentangle the background and growth factor behaviors; their consistency is a key prediction of the so-called smooth dark energy paradigm~\citep{Mortonson:2008qy}. Following~\citet{Zhong:2023how}, the parameters that only modify the growth factor are $\Omega_m^\mathrm{growth}$ and $w^\mathrm{growth}$. Finally, $\Lambda$CDM-split is the particular case where $w^\mathrm{geo} = w^\mathrm{growth} = -1$. 

This paper is organized as follows. In Sec.~\ref{sec:methodology}, we present the assumptions for LSST-Y1 forecasting and the main steps for implementing our proposed emulator. In Sec.~\ref{sec:validation_of_emulator}, we provide details of the emulator training and validation. As an application in an extension to $\Lambda$CDM, we provide LSST-Y1 forecasts for the so-called growth-geometry $w$CDM split in Sec.~\ref{sec:gg-split}. Finally, we discuss the possible use of the emulator and compare it with previous work in Sec.~\ref{sec:conclusion}.

\section{Methodology}\label{sec:methodology}

We detail in this paper the training and validation of machine-learning emulators that can simulate the real-space data vector of cosmic shear ($\xi_{+}, \xi_{-}$), galaxy-galaxy lensing ($\gamma_t$), galaxy clustering ($w$), and their combination (3x2pt) in the context of Rubin Observatory year one (LSST-Y1)~\citep{DESC-SRD}. Our adopted cosmology is the growth-geometry $w$CDM split model~\citep{Wang:2007fsa,Ruiz:2014hma,DES:2020iqt, Zhong:2023how}, which we introduce in Sec.~\ref{sec:gg-split}. Regarding the LSST-Y1 forecast choices, we closely follow the methodologies outlined in~\citet{Boruah:2022uac}, which are themselves based on the DESC collaboration science requirement document~\citep{DESC-SRD}. The first year's data from Rubin is expected to have an observed survey area of $12300 \ \mathrm{deg}^2$, with shape noise of $\sigma_e = 0.26$, and an effective number density of source galaxies $n_\mathrm{eff}=11.2 \, \mathrm{arcmin}^{-2}$. 

The 3x2pt data vector has 1560 data points before scale-cuts, divided into five tomographic lens bins and five tomographic source bins with angular scales between 2.5 and 900 arcmin. In both the training and validation of our emulator, we included all data points without any prior scale cuts. However, we later apply the scale-cut $\theta_\mathrm{min} = \left\{2.756, 8.696\right\}$ arcmin on the cosmic shear components $\left\{\xi_{+}, \xi_{-}\right\}$. We additionally applied angular scale cuts corresponding to the physical scale $R_{\min}=21  h^{-1} \, \mathrm{Mpc}$ when computing galaxy-galaxy lensing. We calculated the covariance matrix with \textsc{CosmoCov}~\citep{Fang2020,Krause:2016jvl}, accounting for the Gaussian, connected non-Gaussian, and super-sample components. 

The systematics modeling incorporates multiplicative shear calibration $m_i$, photometric redshift uncertainties with shifts $\Delta z_{l/s}^i$ (lens and source bins), intrinsic alignment (IA) via the non-linear alignment modeling~\citep{Bridle:2007ft}, and linear galaxy bias $b_i$. These parameters are listed in Table~\ref{table:priorchoices}. We compute the data vectors using the \textsc{Cosmolike} software~\citep{2014MNRAS.440.1379E,Krause:2016jvl}, and manag the MCMC chains using the \textsc{Cobaya} sampler~\citep{Torrado:2020dgo}. Both packages is integrated in the \textsc{Cobaya}-\textsc{Cosmolike} Architecture, or \textsc{CoCoA}\footnote{\url{https://github.com/CosmoLike/cocoa}}. 

\begin{table}
\centering
\renewcommand{\arraystretch}{1.2}
\begin{tabular}{lcc} %
    \hline
    {Parameter} & {Fiducial Value} & {Prior}\\
    \hline
    \textbf{Standard Cosmology}\\
    $\Omega_\mathrm{m}^{\rm geo}$ & $0.3166$ & \textsc{flat}$[0.24 , 0.4]$ \\
    $\Omega_\mathrm{b}$ & $0.0493$ & \textsc{flat}$[0.04 ,  0.06]$ \\
    $\ln (10^{10} A_{\mathrm{s}})$ & $3.045$ & \textsc{flat}$[2.84 ,  3.21]$ \\
    $ n_s $ & $0.9649$ & \textsc{flat}$[0.92, 1.0]$ \\
    $ H_0 $ & $67.32$ & \textsc{flat}$[61, 73]$ \\
    $w^{\rm geo}$ & $-1$ & \textsc{flat}$[-1.3 , -0.7]$ \\
    $ \tau $ & - & \textsc{flat}$[0.01, 0.2]$ \\
    \hline
    \textbf{Split Parameters} \\
    $\Omega_\mathrm{m}^{\rm growth}$ & $0.3166$ & \textsc{flat}$[0.24 , 0.4]$ \\
    $w^{\rm growth}  $ & $-1$ & \textsc{flat}$[-1.3 , -0.7]$ \\
    \hline
    \textbf{Linear Galaxy bias} \\
    $b_1$ & $1.73$ & \textsc{flat}$[0.8 , 3.0]$ \\
    $b_2$ & $1.65$ & \textsc{flat}$[0.8 , 3.0]$ \\
    $b_3$ & $1.61$ & \textsc{flat}$[0.8 , 3.0]$ \\
    $b_4$ & $1.93$ & \textsc{flat}$[0.8 , 3.0]$ \\
    $b_5$ & $2.12$ & \textsc{flat}$[0.8 , 3.0]$ \\
    \hline
    \textbf{\small Intrinsic Alignment} \\
    $a_{1}$    & $0.5$   & \textsc{flat}$[-5 , 5]$ \\
    $\eta_{1}$ & $0$ & \textsc{flat}$[-5 , 5]$ \\
    \hline
    \textbf{Source photo-z} \\
    $\Delta z_{\mathrm{s}}^{i} \times 10^{2}$ & $0$ & \textsc{Gauss}$[0, 0.2]$ \\
    \hline
    \textbf{Lens photo-z}\\
    $\Delta z_{\mathrm{l}}^{i} \times 10^{2}$ & $0$ & \textsc{Gauss}$[0, 0.2]$ \\
    \hline
    \textbf{Shear calibration} \\
    $m_{i} \times 10^2$ & $0$ & \textsc{Gauss}$[0, 0.5]$ \\
    \hline
\end{tabular}
\caption{Prior distributions and fiducial values for the cosmological and nuisance parameters utilized in our LSST-Y1 forecasts. The priors conform to either the uniform distribution, displayed as \textsc{flat[min, max]}, or the normal distribution, displayed as \textsc{Gauss[mean, standard deviation]}. When training the emulators, we set the bounds of the Latin Hypercube and uniform samplings that distributed the parameters to the limits of the flat priors except for $\tau$. We did not expand the training edges to avoid degradation of the emulator performance near the boundaries. We also set hard edges on parameters that obey the normal distribution to $[-3.3 \, \times \, \mathrm{standard} \, \mathrm{deviation}, +3.3 \, \times \, \mathrm{standard} \, \mathrm{deviation}]$ when training the emulator. However, when running the chains, we did not apply these additional cuts.}
\label{table:priorchoices}
\end{table}

\begin{table}
\centering
\renewcommand{\arraystretch}{1.3}
\begin{tabular}{lr} 
\hline
Hyper-parameter & Cosmic Shear (2x2pt) \\
\hline
\textbf{Starting Learning Rate} & $5\times10^{-4}$ \\
\hline
\textbf{Number of Epochs} & $400$ ($600$) \\
\hline
\textbf{Batch Size} & $1500$ \\
\hline
\textbf{\textsc{AdamW}} $\boldsymbol{\beta_1}$ & $0.9$  \\
\hline
\textbf{\textsc{AdamW}} $\boldsymbol \beta_2$ & $0.999$  \\
\hline
\textbf{\textsc{AdamW}} $\boldsymbol \lambda$ & $0.01$  \\
\hline
\end{tabular} 
\caption{Hyper-parameter values we adopted when training the cosmic shear and 2x2pt emulators. We also employed the Adam minimizer with decoupled weight decay, also known as \textsc{AdamW}, following~\citet{2017arXiv171105101L}. The parameters $\beta_1$, $\beta_2$ and $\lambda$ closely follow the nomenclature presented in~\citet{2017arXiv171105101L}.}
\label{table:hyper_parameter}
\end{table}

\subsection{Emulator Design and Training}\label{sec:emulator}

The baseline training set consists of two million (2M) data vectors, and we also present results with enhanced sets of four (4M) and eight (8M) million training points. As our first guess on the required training size, we compute one million data vectors distributed according to a Latin Hypercube Sampling (LHS) (using package \textsc{PyDOE-v0.3.6}). LHS is known for its effective space-filling properties~\citep{2022arXiv220306334D}. However, its distribution varies based on the total number of samples. Combining two sets generated following LHS distributions results in correlated overall training data that does not have the same LHS properties. In this case, we observed a worsening of the emulator performance, so each LHS set must be generated independently. We generate all additional data vectors necessary to assemble the 2M, 4M, and 8M training sets following a uniform distribution. We could then expand the 1M initial sample to 2M, 4M, and 8M without recomputing the entire set from scratch. We did not further explore other ways of constructing data sets or other space-filling sampling methods such as the Sobol sequence~\citep{SOBOL196786}.

Given the prior knowledge of training parameters, we were able to employ both shared-memory (via OpenMP)~\citep{dagum1998openmp} and message-passing (via MPI)~\citep{10.5555/898758} parallelization schemes when calculating the training data vectors. We compute the data vectors at the approximate rate of 15,000 data vectors per hour per 40-core node. Our computational nodes are powered by two \textsc{Intel Xeon Gold 6148}, delivering approximately 2,168 million operations per second per core\footnote{\url{https://www.cpubenchmark.net/cpu.php?cpu=Intel+Xeon+Gold+6148+\%40+2.40GHz&id=3176}}. 
The quoted numbers illustrate that training can be accomplished in less than a day with moderate resources and computational costs of the order of a few thousand CPU hours. 

\begin{figure*}
\centering
\includegraphics[width=2.0\columnwidth]{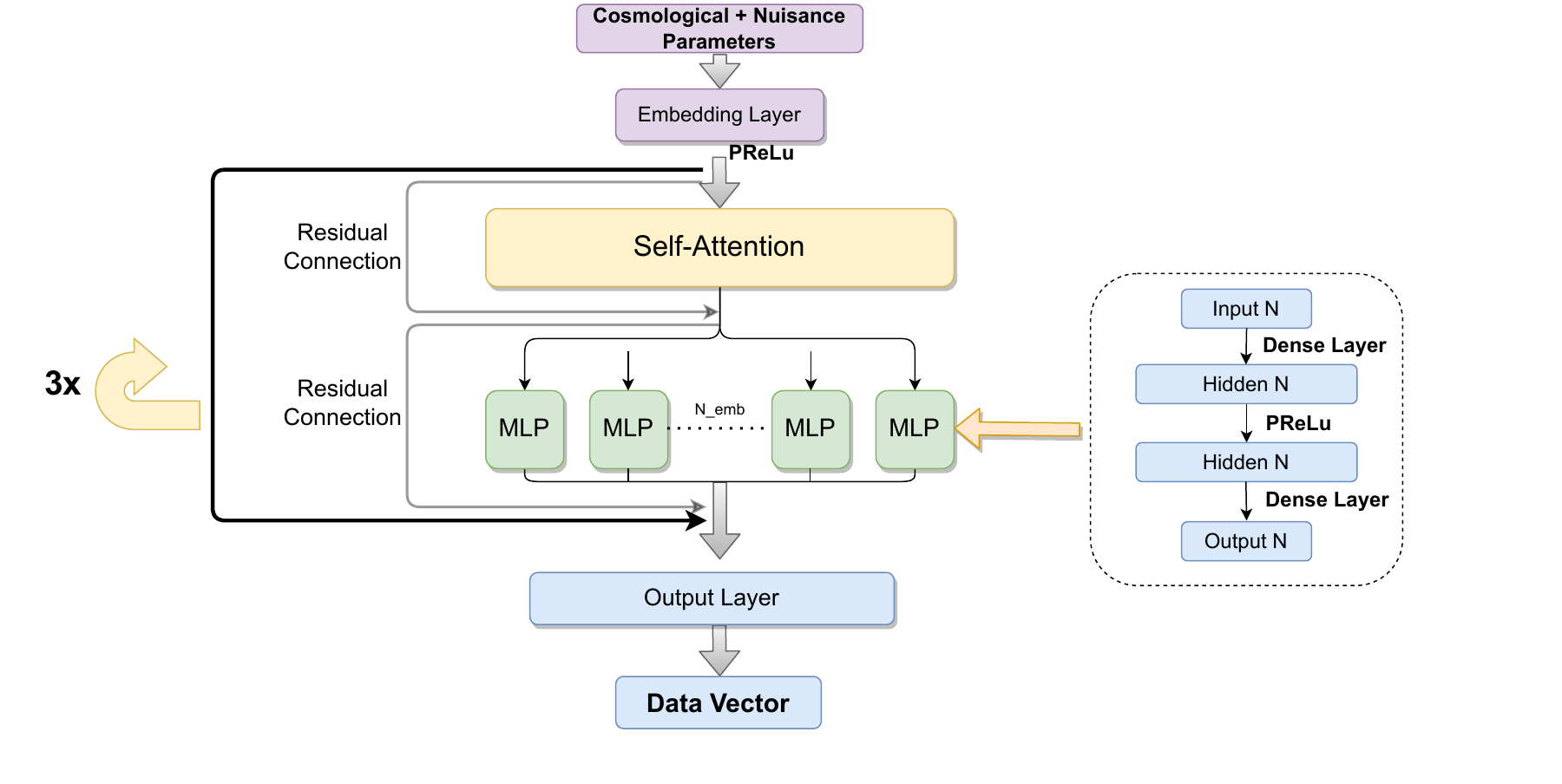}
\caption{Proposed transformers-based neural network architecture. The input array has a dimension of 13, representing the growth-geometry split $\Lambda$CDM parameters plus the cosmic shear nuisance parameters. In the embedding layer, we employ $N_\mathrm{emb} = 25$ linear transformations that take in the cosmological and nuisance parameters and produce output vectors with dimension 200. We set an activation function \textsc{PReLU}~\citep{2015arXiv150201852H} between the embedding layer and the transformer blocks. We then placed three transformer blocks. Each transformer block has a self-attention layer and $N_\mathrm{emb}$ parallel fully connected multilayer perceptrons (MLP). We employ normalization layers after the self-attention and MLP blocks~\citep{2020arXiv200307845S} to ensure proper normalization of the intermediate output products. We add a residual connection (the identity operator) that adds the input vectors to the output of the self-attention block. We also added a residual connection that adds the input vector to the output of the MLP block. The MLP block, magnified in the figure above, deserves a more detailed explanation. In a few known implementations of the transformer block, the output first linear transformation is a vector with dimension $\tilde{N}$ (drawn as {\it hidden N}), which differs from the input dimension N. In cases where $N \neq \tilde{N}$, there is a second linear transformation after the \textsc{PReLU} activation function so that the output vector also has dimension $N$. We tried different $(\tilde{N}, N)$ combinations but found that $\tilde{N} = N$ offers good performance. In this case, the presence of the second linear layer is redundant, but we kept it to maintain the generality of the design. We also follow the original transformer design to keep the multiple MLPs the same within one block, this weight-sharing reduces number of trainable parameters in the architecture, which improved the accuracy in our settings. Finally, the output layer output vectors with dimension $780$, correspond to the size of the cosmic shear data vector. For 2x2pt emulators, we adjusted the dimensionality of the input and output layers accordingly.} 
\label{fig:nn_design} 
\end{figure*}

Our training effectively covers the entire parameter space adopted on the growth-geometry split study~\citet{Zhong:2023how}. In this work, the cosmological parameters are restricted to a prior analogous to the one adopted by the Euclid collaboration when creating version 2.0 of the \textsc{Euclid Emulator} for the matter power spectrum~\citet{Euclid:2020rfv}. We validate the performance of all emulators using a loss function that equally weighs samples in the entire prior, being attentive in ensuring their accuracy was not position-dependent. While the \textsc{Euclid Emulator} prior is still informative on MCMC chains containing only LSST-Y1 likelihood, the volume it encompasses is much larger than what previous emulators could simulate accurately~\citep{Boruah:2022uac, 2023JCAP...01..016T}.

We implement the neural networks with \textsc{PyTorch}~\citep{pytorch}, and adopt the \textsc{AdamW} as the gradient-based optimizer~\citep{2014arXiv1412.6980K, 2017arXiv171105101L} when training the machine learning models. Table~\ref{table:hyper_parameter} shows the starting learning rate, batch size, and number of epochs utilized in training. These values reflect our experience on what constitutes, loosely speaking, a good solution after several rounds of training and validation. Therefore, different applications may require further fine-tuning. We compare three neural network designs: a fully connected multilayer perceptron (\textsc{MLP}\footnote{MLP is a naming convention for feedforward neural networks with a single input and output, consisting of a stack of dense layers separated by activation functions and without recurring or residual connections.}), a 1D residual network (\textsc{ResMLP}\footnote{In the literature, the term \textsc{ResMLP} is sometimes referred to as \textsc{ResNet}. Both terms can be confusing as they appear in many contexts to denote wildly different implementations of neural networks. In this work, \textsc{ResMLP} is used to denote a feedforward neural network with a single input and output, consisting of a stack of dense layers separated by activation functions and with residual connections connecting every two layers.}), and our proposed transformers-based network (TRFB) is described in Fig.~\ref{fig:nn_design}. 

We built the MLP and \textsc{ResMLP} emulators loosely following the prescriptions described in~\cite{Boruah:2022uac} and~\cite{2023JCAP...01..016T}. However, this paper should not be interpreted as providing a direct comparison between our proposed TRFB and their implementations; a comparison would require all strategies to be fully optimized. The MLP design consists of five dense layers stacked together, each of size 512. We included the \textsc{PReLU} activation function between the layers~\citep{2018arXiv180308375A}. The \textsc{ResMLP} design has a similar set of stacked layers, but we increase their size to 1024. \textsc{ResMLP} also contains residual connections between every two layers, which adds the original vector from the top layer to the output of the bottom one. 

The training time, not counting what was needed to compute the data vectors, varied considerably depending on the pipeline design and hyper-parameter choices. Typically, it is around five hours for \textsc{MLP} and \textsc{ResMLP} models, whereas the attention-based network requires about ten hours and significantly more memory. This increased memory consumption led us to employ the high-end A100 GPU for training. Lower-tier GPUs with less memory could be used as well if, for instance, the batch size is reduced, but we have not tested how this would impact training time. We only stress that without further optimizations, Transformers is more memory-intensive than \textsc{MLP} and \textsc{ResMLP}. 

We trained the cosmic shear and galaxy-galaxy lensing data vectors separately, as this increased the validation accuracy considerably. Cosmic shear and 2x2pt have similar output lengths but different input parameters; the 2x2pt has five additional nuisance parameters related to the lens's photo-z bias and is thus more challenging to train. We did not observe additional enhancements in the final accuracy by further decomposing the data vector into smaller pieces. 

\subsubsection{Emulator Input and Output}

The input parameters, $\boldsymbol{\Theta}$, include both cosmological and nuisance parameters, which vary in their values by orders of magnitude and have different prior ranges $[\Theta_{i,\min},\Theta_{i,\max}]$. We then normalize them via the redefinition
\begin{equation}
\boldsymbol{\theta}_i=\frac{\Theta_i-\Theta_{i,\min }}{\Theta_{i,\max }-\Theta_{i,\min}} \, ,
\end{equation}
so all components of the vector $\boldsymbol{\theta}$ are between $[0,1]$. 

We match the training set boundaries to the uniform priors displayed in Table~\ref{table:priorchoices}, except for three modifications. First, we apply an additional cut in the directions of parameter space with Gaussian prior, limiting training points to the interval $[-3.3 \, \times \, \mathrm{standard} \, \mathrm{deviation}, +3.3 \, \times \, \mathrm{standard} \, \mathrm{deviation}]$. Second, we do not include the reionization optical depth $\tau$ as it is not a model parameter for weak lensing data vectors. Finally, we exclude shear multiplicative and galaxy linear biases since their effects can be computed with a simple multiplication as shown below (they are fast parameters in MCMC chains)
\begin{equation*}
\begin{aligned}
\xi_{\pm}^{i j}(m_{i},m_{j}) & \rightarrow \left(\frac{1+m_i}{1+m_i^\mathrm{fid}}\frac{1+m_j}{{1+m_j^\mathrm{fid}}}\right) \xi_{ \pm}^{i j}(m_i = m_i^\mathrm{fid},m_j = m_j^\mathrm{fid}),
\end{aligned}
\end{equation*}
\begin{equation*}
\begin{aligned}
\gamma_t^{ij}(m_s) & \rightarrow\left(\frac{1+m_j}{{1+m_j^\mathrm{fid}}}\right) \gamma_t^{ij}(m_j=m_j^\mathrm{fid}),
\end{aligned}
\end{equation*}
\begin{equation}\label{eq:fast_parameters}
\begin{aligned}
\gamma_t^{ij}(b_i) & \rightarrow \left(\frac{b_{i}}{b_{i}^\mathrm{fid}} \right)\gamma_t^{ij}(b_i=b_{i}^\mathrm{fid}), 
\end{aligned}
\end{equation}
The indexes represent redshift bins. This exclusion reduces the input dimension and simplifies the training process. For the purpose of forecasting, we use Limber approximation and do not consider redshift spatial distortion (RSD) or lensing magnification. Including these two systematics or performing the full non-Limber calculation will make the bias parameter scaling relation not linear. Another drawback of the parametrization as in Eq.~\ref{eq:fast_parameters} is that it decreases the emulator's precision whenever the fast parameters are much larger than the fiducial values even when the linear scaling relation holds. This is because the error propagates when it is being multiplied by the true $1+m_i$ or $b_i$. However, since we impose a stringent test for the precision of the neural network and sample fast parameters within the same order as the fiducial value, we do not expect this to be a problem. In the future, we plan to include bias parameters in the direct emulation.

\begin{figure}
\centering
\includegraphics[width=\columnwidth]{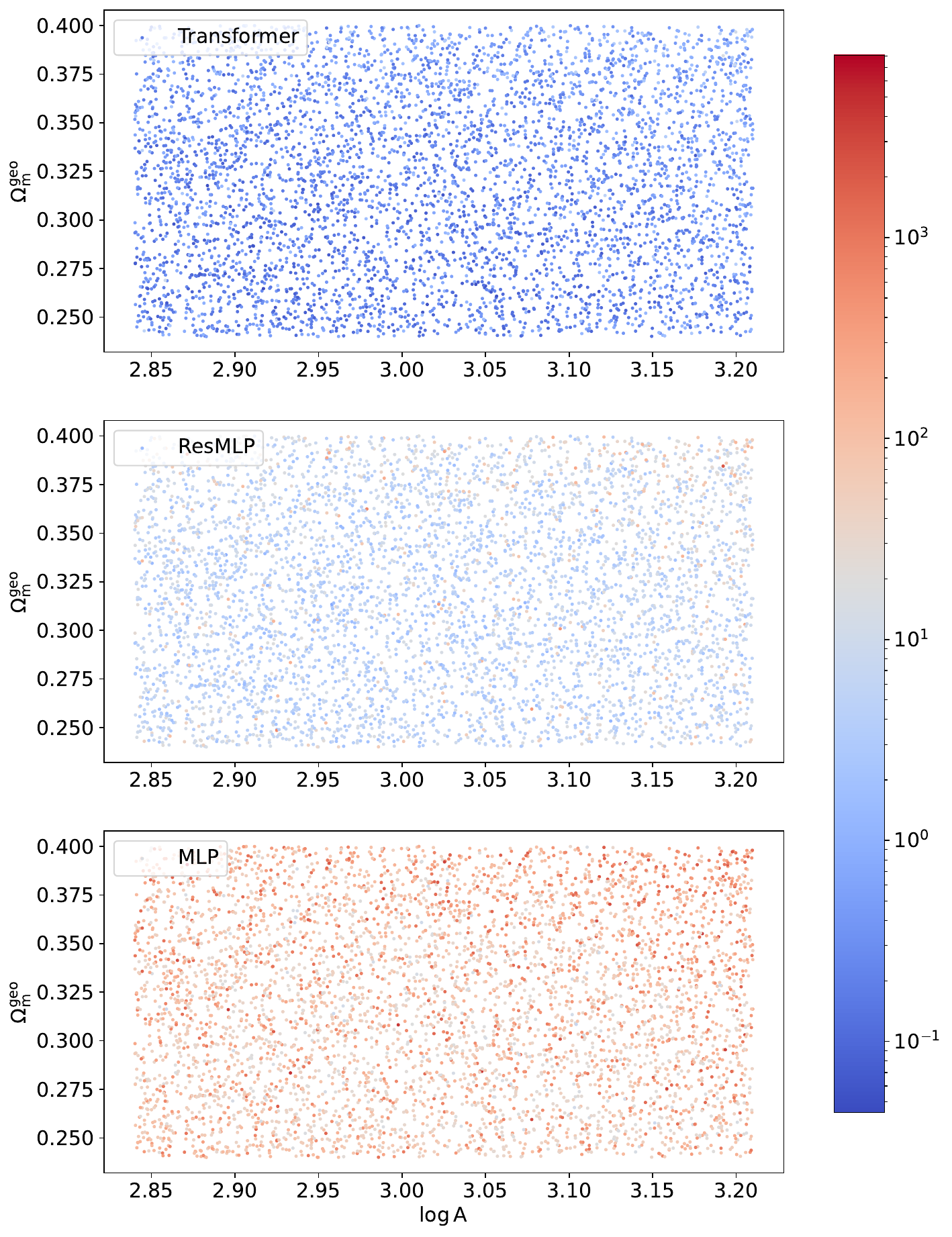}
\caption{Projected spatial distribution of the $\Delta \chi^2$ residuals, computed on validation points, between the emulators and \textsc{CoCoA}, in the two-dimensional $(\Omega_m^\mathrm{geo},\log(A_s))$ plane. The figure highlights the performance difference between the transformers-based TRFB (top panel), \textsc{RestNet} (middle panel), and MLP (bottom panel) designs. The training was limited to the cosmic shear data vector, but it involved varying all growth-geometry $\Lambda$CDM split and nuisance parameters, except for the shear multiplicative biases. All emulators were trained with the baseline two million (2M) data vectors and tuned with the same starting learning rate, number of epochs, and batch size hyper-parameters. After applying the scale cut $\theta_\mathrm{min} = \{2.756, 8.696\}$ arcmin for $\left\{\xi_{+}, \xi_{-}\right\}$, the average (median) $\Delta \chi^2$ values were 0.52 (0.30) for TRFB, 13.0 (6.1) for \textsc{ResMLP}, and  221.8 (76.5) for MLP. The mean and median differ by a factor of two due to significant degradation in $\Delta \chi^2$ when parameters are very close to the prior edge. For the TRFB emulator, only $10\%$ of the validation points exhibited $\Delta \chi^2>1$, the adopted DES-Y3 threshold for errors to require correction via importance sampling to prevent biasing the goodness of fit.}
\label{fig:nn_validation} 
\end{figure}

Instead of directly outputting the cosmic shear or the 3x2pt data vectors, the emulator outputs a related normalized vector, denoted $\boldsymbol{\mathcal{D}}$. To transform $\boldsymbol{y} \equiv \{\xi_{+}, \xi_{-}\}$ or  $\{\gamma_t, w \}$ into $\boldsymbol{\mathcal{D}}$, we first apply the change of basis
\begin{equation}
\boldsymbol{\Bar{y}} = \boldsymbol{y} Q \, ,
\end{equation}
where the matrix $Q$ contain the eigenvectors of the cosmic shear or 3x2pt covariance matrix, $\Sigma$, as shown below
\begin{equation}
\mathbf{\Sigma} \equiv Q \Lambda Q^{-1} \, . 
\end{equation}
Here, it is possible to disregard eigenvector directions (principal components) associated with large errors; they are mostly irrelevant as the data cannot adequately constrain them. We kept all the principal components in this study for simplicity (one less parameter to be optimized). Finally, we apply the normalization 
\begin{equation}
\boldsymbol{\mathcal{D}}=\frac{\boldsymbol{\Bar{y}}-\boldsymbol{\Bar{y}}_\mathrm{mean}}{\sigma_y} \, .
\end{equation}
where $\boldsymbol{\Bar{y}}_{\mathrm{mean}}$ is the mean of the data vectors and $\sigma_y=\sqrt{\operatorname{diag}[\Lambda]}$. We stored the change of basis and normalization factors alongside the neural network, enabling us to transform back to the original data vector.

\begin{figure}
\centering
\includegraphics[width=0.94\columnwidth]{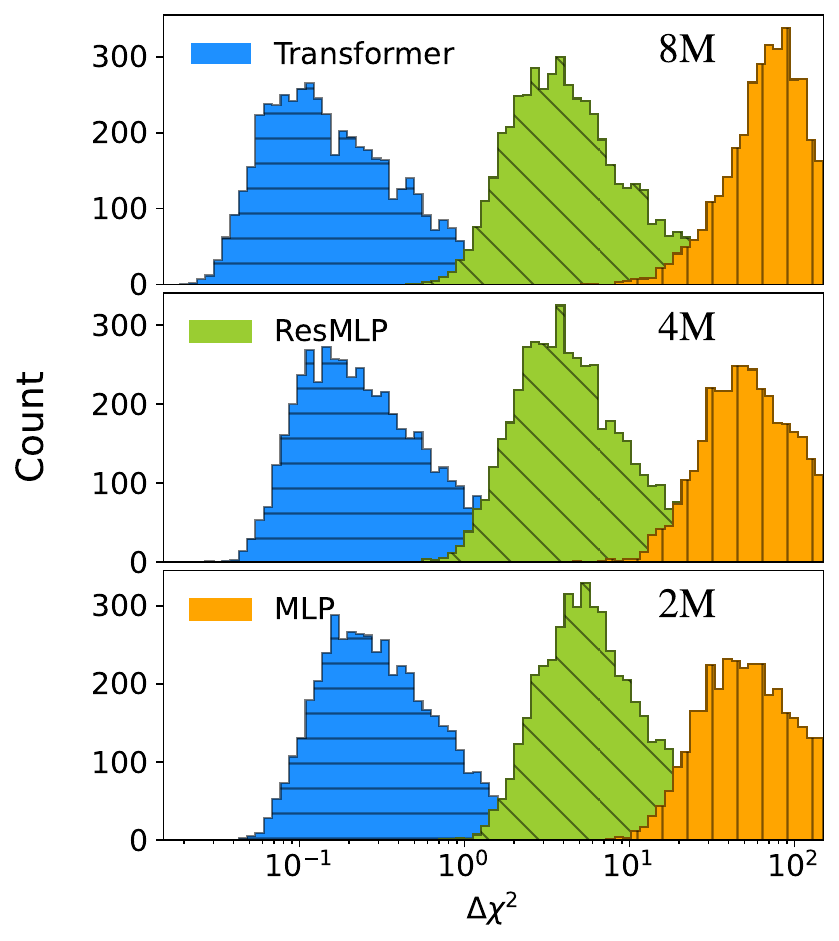}
\caption{The cosmic shear $\Delta \chi^2$ residuals between three emulators and \textsc{CoCoA}, calculated on the validation set. The transformers-based TRFB emulator is shown on the blue histogram (with horizontal lines), \textsc{ResMLP} on the green histogram (with diagonal lines), and MLP on the orange histogram (with vertical lines). We trained all emulators with the same starting learning rate, number of epochs, and batch size hyper-parameters, detailed in Tab.~\ref{table:hyper_parameter}. Although additional fine-tuning of these hyper-parameters could improve the \textsc{ResMLP} and MLP residuals, the enormous impact of the neural network architecture is evident. The bottom, middle, and top panels demonstrate the limited influence of increasing the size of the training set from two (2M) to eight million (8M) points. When the TRFB emulator is trained with 8M points, only $4\%$ of the validation points have $\Delta \chi^2 > 1$, the adopted DES-Y3 threshold
for errors to require corrections via importance sampling to prevent biasing
the goodness of fit. Additionally, only $38.1\%$ of the validation points exhibit residuals larger than $\Delta \chi^2 = 0.2$, the adopted DES-Y3 threshold for a model variation without refitting to be considered insignificant~\citep{DES:2021rex}.} 
\label{fig:chi2_list} 
\end{figure}

\subsubsection{Loss Function}\label{sec:loss_function}

We define the loss function to be the average $\Delta\chi^2$ between the data vectors computed using the neural network, $\boldsymbol{y}_{\mathrm{NN}}$, and those calculated from \text{CoCoA}, $\boldsymbol{y}_{\mathrm{CoCoA}}$. The average intentionally excludes any positional-dependent weight that could emphasize the accuracy near or far from the center of the prior volume, as shown below.
\begin{equation}
L(\boldsymbol{w}) = \left< \left( \boldsymbol{y}_{\mathrm{NN}} - \boldsymbol{y}_{\mathrm{CoCoA}} \right)^{T} \Sigma^{-1} \left( \boldsymbol{y}_{\mathrm{NN}} - \boldsymbol{y}_{\mathrm{CoCoA}} \right)  \right> \, .
\end{equation}
Here, $\boldsymbol{w}$ represents the trainable weights in the machine learning design. This approach balances the emulator accuracy across the entire prior volume and guides the training to focus on the redshift and angular bins with a smaller error bar.

Furthermore, the average $\Delta\chi^2$ provides a good (but incomplete) indication of whether analyses utilizing emulators need the data likelihoods to be corrected via the Bayesian error propagation proposed in~\citet{2022arXiv220511587G}. The drawback of this loss choice is its sensitivity to outliers; therefore, we show the entire $\Delta \chi^2$ distribution on validation points in Section~\ref{sec:validation_of_emulator}. We left for future work investigations on alternative schemes such as the ranking and elimination of the worst few percent of points (the exact threshold to be fine-tuned) when computing the average $\Delta\chi^2$ in both training and validation.

\subsubsection{Architecture}\label{sec:transformer}

We adopt a neural network design centered on the Transformer architecture that employs the self-attention mechanism~\citep{2014arXiv1409.0473B,Attention_is_all_you_need}. Transformers and its underlying self-attention have been used successfully in the construction of large language models with many potential applications in physics and astronomy~\citep{2023arXiv231006083B, 2023RAA....23j4005C, 2023arXiv231114972C, 2023SpWea..2103485C, 2023JHEP...06..184F, leung2023towards, 2023arXiv230612384L, 2023JHyd..62530085N, 2023arXiv230916316P, 2023Geop...88C.107S,2023FrP....1166927Z}. 

\begin{figure}
\centering
\includegraphics[width=0.99\columnwidth]{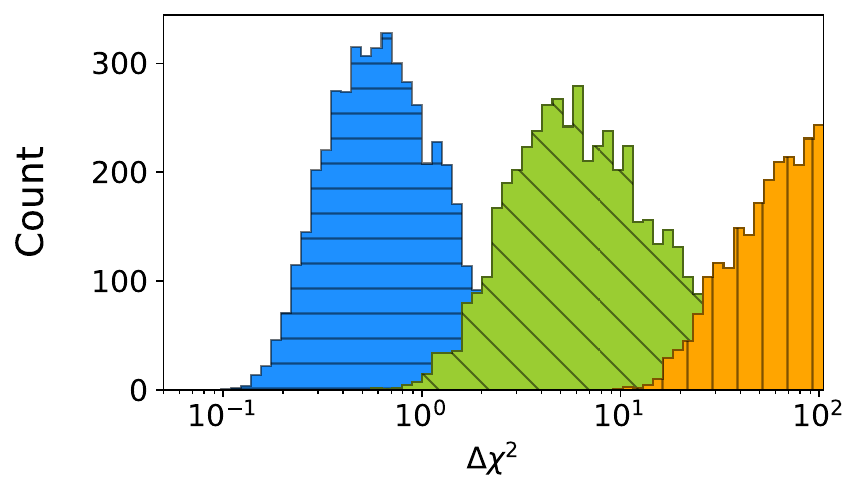}
\caption{The 3x2pt $\Delta \chi^2$ residuals between the emulators and \textsc{CoCoA}, calculated on the validation set. The transformers-based emulator is shown on the blue histogram (with horizontal lines), \textsc{ResMLP} on the green histogram (with diagonal lines), and MLP on the orange histogram (with vertical lines). In all three cases, we independently emulated the cosmic shear and the galaxy-galaxy lensing data vectors. We trained all emulators with the same starting learning rate, number of epochs, and batch size hyper-parameters, detailed in Table.~\ref{table:hyper_parameter}. We trained the 2x2pt emulators with two million points, and use the cosmic shear emulator that we have trained with eight million points.}
\label{fig:chi2_list2} 
\end{figure}

\begin{figure*}
\centering
\includegraphics[width=\columnwidth]{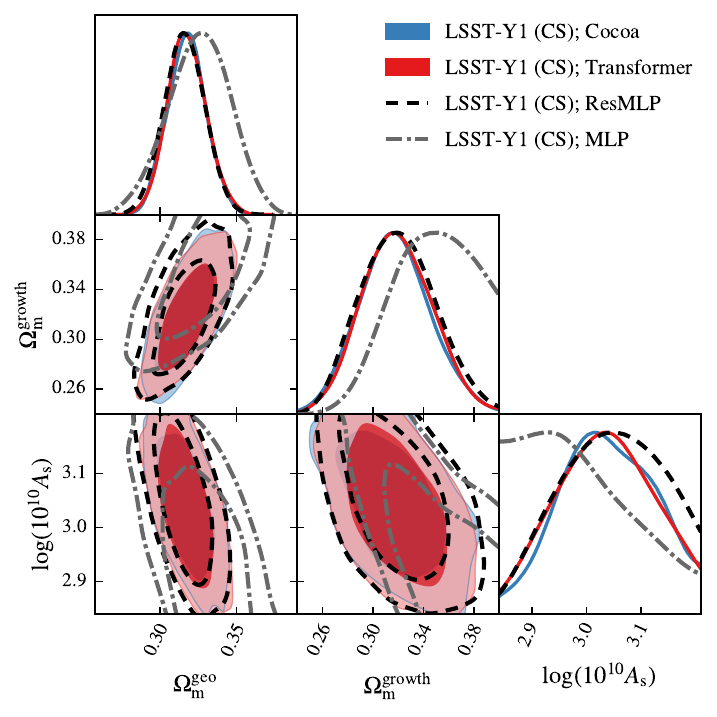}
\includegraphics[width=\columnwidth]{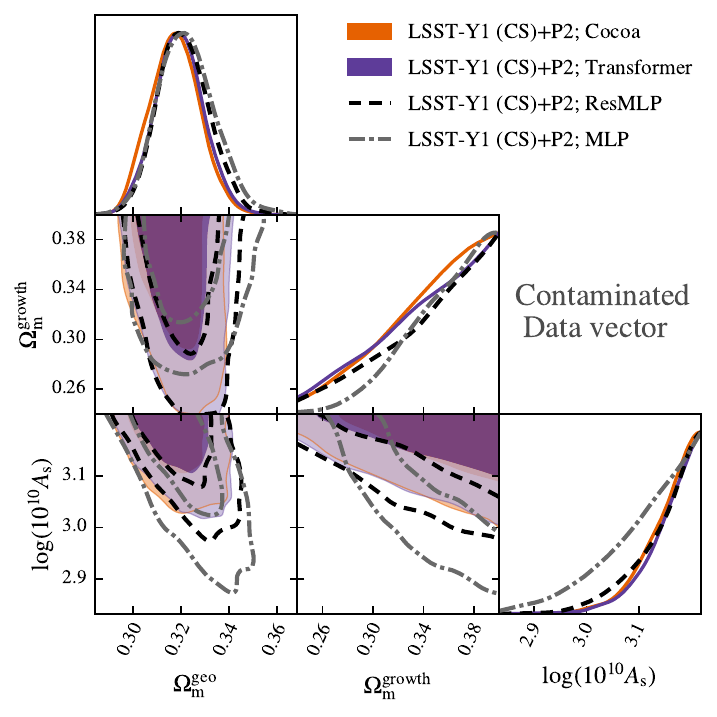}
\caption{Posterior distributions when employing either machine learning emulators or \textsc{CoCoA} to compute the cosmic shear data vector. We computed the fiducial data vector adopted on all MCMC chains with \textsc{CoCoA}. In the left panel, the fiducial cosmology is the $\Lambda$CDM model with values displayed in Tab.~\ref{table:priorchoices}, and the scales cut are $\theta_\mathrm{min} = {2.756, 8.696}$ arcmin for $\left\{\xi_{+}, \xi_{-}\right\}$ respectively. The right panel's fiducial model includes a dark energy component with anisotropic stresses, further elaborated in Appendix~\ref{sec:ppf}. In this case, we made more conservative scale cuts to ensure that information from nonlinear scales is irrelevant. The exotic fiducial dark energy shifted the posterior distribution of many cosmological parameters. Compared to the left panel, this case provides a more stringent test on the consistency between the machine learning emulators and \textsc{CoCoA} far from the center of the prior. These chains are also illustrative of the usefulness of the growth-geometry $\Lambda$CDM split as an indicator of cracks in the validity of the smooth dark energy paradigm. Consistent with the excellent agreement we observed on $\Delta \chi^2$ tests, there is no appreciable difference in the posteriors between the transformers-based emulator and \textsc{CoCoA}. The \textsc{ResMLP} emulator also did not generate biased posteriors, but there is a visible worsening in the agreement to \textsc{CoCoA}. The MLP emulator exhibits the poorest performance.}
\label{fig:posterior_validation} 
\end{figure*}

The transformer architecture excels at processing sequential data, such as language and time series. Its key innovation lies in the self-attention mechanism, which calculates a weighted sum of a sequence of input vectors~\citep{2015arXiv150203044X,2016arXiv160106733C}. These weights are computed from mathematical operations involving three types of vectors: a query $\mathbf{q}$, a key $\mathbf{k}$, and a value $\mathbf{v}$. In addition, the particular operation between the query, the key, and the value vector defines different attention mechanisms. The relation between the input sequence of data and $\{Q, K, V\}$ (series of queries, keys, and values) are given by the projection matrices $\{W_Q, W_K, W_V\}$ with trainable parameters. The scaled dot-product attention is then defined via the operation:

\begin{equation}
\operatorname{Attention}(Q, K, V)=\operatorname{softmax}\left(\frac{1}{\sqrt{d_k}} Q K^T\right) V,
\end{equation}
where $d_k$ is the dimension of the key vector. The self-attention mechanism allows the transformer model to process data more efficiently because it can capture long-range dependencies between elements of the sequence without requiring the computation of a large number of sequential operations (for instance, a long stack of dense layers). In our case, there are long-range similarities between each tomographic bin. See~\citet{2023arXiv231012069T} for a more in-depth review.

The complete design is illustrated in Fig.~\ref{fig:nn_design}. The size of the total trainable parameters is about five million. The first embedding layer expands the sixteen input parameters into a set of twenty-five vectors of dimension two hundred. This process essentially maps the input data into a higher-dimensional space. These vectors serve as a richer representation of the input features, capturing complex relationships and patterns that might not be apparent in the original parameters.  After the self-attention layer and the multi-head MLP layers, we applied the layer norm~\citep{2020arXiv200307845S}, an alternative to the conventional batch norm. Layer norm is more commonly used in transformer models~\citep{Attention_is_all_you_need}, but we did not find a difference between using the two normalization schemes. The transformer block also contains residual connections that help to reduce the so-called degradation problem~\citep{2015arXiv151203385H}. Finally, we did not add positional embeddings into the vectors~\citep{Attention_is_all_you_need}.

\section{Emulator Validation}\label{sec:validation_of_emulator}

We validate the emulators against \textsc{CoCoA} with five thousand random points uniformly sampled from the prior. Fig.~\ref{fig:nn_validation} compares the TRFB, the \textsc{ResMLP}, and the MLP emulators on computing cosmic shear data vectors. All cosmic shear emulators are trained with two million data vectors, which we defined as the baseline scenario. We don't introduce any noise in the data vectors, so we expect zero mean and median $\Delta \chi^2$ in the validation set for a perfectly accurate emulator. The mean (median) $\Delta \chi^2$ between the emulators and \textsc{CoCoA} prior to applying any scale cuts are
$\langle \Delta \chi^2_\mathrm{TFRB} \rangle = 0.65 \, (0.38)$, $\langle \Delta \chi^2_\mathrm{ResMLP} \rangle = 17.0 \, (7.6)$, and $\langle \Delta \chi^2_\mathrm{MLP} \rangle = 267.8 \, (101.4)$. All panels display the  $\Delta \chi^2$ after the scale cuts have been applied, and in this case $\langle \Delta \chi^2_\mathrm{TFRB} \rangle = 0.52 \, (0.30)$, $\langle \Delta \chi^2_\mathrm{ResMLP} \rangle = 13.0 \, (6.1)$ and $\langle \Delta \chi^2_\mathrm{MLP} \rangle = 221.8 \, (76.5)$. 

Figure.~\ref{fig:nn_validation} shows, for all three emulators, a uniform validation distribution of $\Delta \chi^2$ values across the two-dimensional projected $\{\Omega_m^\mathrm{geo}, \log(A_s) \}$ plane. We did not detect significant $\Delta \chi^2$ spatial variations except close to the prior boundaries. This consistency is an important indication that our emulator is accurate in MCMC and nested sampling calculations regardless of the location of the fiducial data vector. Similar conclusions hold for the 3x2pt emulator.

Figure.~\ref{fig:chi2_list} displays the $\Delta \chi^2$ validation distribution for the TFRB (blue), \textsc{ResMLP} (green) and MLP (orange) designs after the cosmic shear scale cuts have been applied. The TRFB emulator improves the overall fit to the \textsc{CoCoA} calculations by one order of magnitude compared to \textsc{ResMLP}. The result is a validation distribution where only a small fraction of points has $\Delta \chi^2 > 1$, the threshold adopted by the Dark Energy Survey (DES) collaboration to indicate an error that could bias goodness-of-fit tests if not approximately corrected via Importance Sampling~\citep{DES:2021rex}.

\begin{table*}
\centering
\setlength{\tabcolsep}{3pt}
\renewcommand{\arraystretch}{1.2}
\begin{tabular}{|ccccccccccc|}
\hline
\multicolumn{11}{|l|}{\textbf{Fiducial Cosmology}}                                                                                                                                                                                                  \\ \hline \hline
\multicolumn{1}{|c|}{$n_\mathrm{live/repeat}$} & \multicolumn{1}{c|}{\small \textsc{CoCoA}}                                               & \multicolumn{1}{c|}{\small TRFB 2M}                                                   & \multicolumn{1}{c|}{\small TRFB 4M}                                                   & \multicolumn{1}{c|}{\small TRFB 8M}                                                & \multicolumn{1}{c|}{\small \textsc{ResMLP} 2M}                                               & \multicolumn{1}{c|}{\small \textsc{ResMLP} 4M}                                               & \multicolumn{1}{c|}{\small \textsc{ResMLP} 8M}                                            & \multicolumn{1}{c|}{\small MLP 2M}                                                  & \multicolumn{1}{c|}{\small MLP 4M}                                                  & \small MLP 8M                                               \\ \hline

\multicolumn{1}{|c|}{1024/5d}   & \multicolumn{1}{c|}{\begin{tabular}[c]{@{}c@{}}-15.81\\ ±0.122\end{tabular}} & \multicolumn{1}{c|}{\begin{tabular}[c]{@{}c@{}}-15.92\\ ±0.122\end{tabular}} & \multicolumn{1}{c|}{\begin{tabular}[c]{@{}c@{}}-15.89\\ ±0.123\end{tabular}} & \multicolumn{1}{c|}{\begin{tabular}[c]{@{}c@{}}-15.85\\ ±0.122\end{tabular}}  & \multicolumn{1}{c|}{\begin{tabular}[c]{@{}c@{}}-15.80\\ ±0.121\end{tabular}} & \multicolumn{1}{c|}{\begin{tabular}[c]{@{}c@{}}-15.81\\ ±0.121\end{tabular}} & \multicolumn{1}{c|}{\begin{tabular}[c]{@{}c@{}}-15.57\\ ±0.121\end{tabular}} & \multicolumn{1}{c|}{\begin{tabular}[c]{@{}c@{}}-16.73\\ ±0.118\end{tabular}} & \multicolumn{1}{c|}{\begin{tabular}[c]{@{}c@{}}-14.89\\ ±0.114\end{tabular}} & \begin{tabular}[c]{@{}c@{}}-16.26\\ ±0.120\end{tabular} \\ \hline \hline
\multicolumn{11}{|l|}{\textbf{Shifted Cosmology 1}}                          
\\ \hline
\multicolumn{1}{|c|}{1024/5d}   & \multicolumn{1}{c|}{\begin{tabular}[c]{@{}c@{}}-15.00\\ ±0.119\end{tabular}}    & \multicolumn{1}{c|}{\begin{tabular}[c]{@{}c@{}}-15.14\\ ±0.120\end{tabular}} & \multicolumn{1}{c|}{\begin{tabular}[c]{@{}c@{}}-15.16\\ ±0.119\end{tabular}} & \multicolumn{1}{c|}{\begin{tabular}[c]{@{}c@{}}-14.97\\ ±0.118\end{tabular}} & \multicolumn{1}{c|}{\begin{tabular}[c]{@{}c@{}}-15.25\\ ±0.118\end{tabular}} & \multicolumn{1}{c|}{\begin{tabular}[c]{@{}c@{}}-15.21\\ ±0.119\end{tabular}} & \multicolumn{1}{c|}{\begin{tabular}[c]{@{}c@{}}-15.07\\ ±0.118\end{tabular}} & \multicolumn{1}{c|}{\begin{tabular}[c]{@{}c@{}}-15.46\\ ±0.115\end{tabular}} & \multicolumn{1}{c|}{\begin{tabular}[c]{@{}c@{}}-14.02\\ ±0.112\end{tabular}} & \begin{tabular}[c]{@{}c@{}}-15.33\\ ±0.113\end{tabular} \\ \hline \hline
\multicolumn{11}{|l|}{\textbf{Shifted Cosmology 2}}                          %
\\ \hline
\multicolumn{1}{|c|}{1024/5d}   & \multicolumn{1}{c|}{\begin{tabular}[c]{@{}c@{}}-13.54\\ ±0.113\end{tabular}}    & \multicolumn{1}{c|}{\begin{tabular}[c]{@{}c@{}}-13.50\\ ±0.112\end{tabular}} & \multicolumn{1}{c|}{\begin{tabular}[c]{@{}c@{}}-13.42\\ ±0.112\end{tabular}} & \multicolumn{1}{c|}{\begin{tabular}[c]{@{}c@{}}-13.53\\ ±0.118\end{tabular}} & \multicolumn{1}{c|}{\begin{tabular}[c]{@{}c@{}}-14.19\\ ±0.111\end{tabular}} & \multicolumn{1}{c|}{\begin{tabular}[c]{@{}c@{}}-13.87\\ ±0.111\end{tabular}} & \multicolumn{1}{c|}{\begin{tabular}[c]{@{}c@{}}-13.84\\ ±0.111\end{tabular}} & \multicolumn{1}{c|}{\begin{tabular}[c]{@{}c@{}}-15.33\\ ±0.109\end{tabular}} & \multicolumn{1}{c|}{\begin{tabular}[c]{@{}c@{}}-13.78\\ ±0.104\end{tabular}} & \begin{tabular}[c]{@{}c@{}}-14.19\\ ±0.106\end{tabular} \\ \hline
\end{tabular}
\caption{Detailed comparison of Bayesian evidence, calculated using the \textsc{PolyChord} nested sampler with hyperparameters set at $n_\mathrm{live}$ = 1024, \textsc{precision criterion} $= 0.001$, and $n_\mathrm{repeats} = 5d$ in all cases. Here, $d$ represents the number of sampled parameters. The numerical results displayed across different columns showcase the cases when we evaluate the cosmic shear data vectors using either \textsc{CoCoA} or the machine learning emulators (TRFB being the transformers-based design). We explored how training set sizes impact Bayesian evidence accuracy in each design, considering three scenarios: training with 2, 4, or 8 million points. We made a similar comparison with shifted data vectors, centered at $(\Omega_\mathrm{m}, \log(A_s)) = (0.287, 3.145)$ and $(\Omega_\mathrm{m}, \log(A_s)) = (0.25, 3.2)$, to assess the impact of prior edge effects on emulator accuracy.}
\label{table:bayesian_evidence}
\end{table*}

Figure.~\ref{fig:chi2_list} also shows that the $\Delta \chi^2$ agreement can be slightly improved in all designs by doubling or quadrupling the points in the training set. However, even eight million (8M) data vectors do not enable the \textsc{ResMLP} architecture to perform better than the baseline TFRB emulator trained with two million (2M) samples. After more than one million points, the design choice for the neural network seems far more important than the training size in determining the final accuracy. This fact does not imply that more data vectors are not desirable. Training the TRFB emulator with 8M points in cosmic shear lowers the median $\Delta \chi^2$ to $0.14$, below the $\Delta \chi^2 = 0.2$ threshold adopted by the Dark Energy Survey year three analysis for a model variation without refitting to be considered insignificant~\citep{DES:2021rex}. We have not tested how far we can push the number of points to have $\Delta \chi^2 < 0.2$. However, comparing the validation distribution of the TFRB emulator trained with 2M versus 8M points suggests that learning has not yet plateaued.

Fig.~\ref{fig:chi2_list2} extends the same conclusion to the 3x2pt data vector emulation with a few modifications. First, we adopt the cosmic shear emulator with four million training points as the baseline when validating the 3x2pt data vectors. Finally, we set the 2x2pt baseline training to contain two and a half million points. These changes reduced the fraction of validation points with $\Delta \chi^2 < 1$ to be similar to what we observed when validating the baseline cosmic shear emulator (with the covariance matrix in the loss function restricted to only the cosmic shear component). We reasonably speculate from Figs.~\ref{fig:chi2_list} and~\ref{fig:chi2_list2} that pushing the number of training points toward the twenty million range for cosmic shear and 2x2pt would drive more than half of validation points to have $\Delta \chi^2 \lesssim 0.2$. 

The compatibility of the TRFB emulator with the $\Delta \chi^2 < 0.2$ requirement would allow it to be used with real data by large collaborations, including the upcoming Dark Energy Survey year six release, without the need for any additional post-processing in the MCMC chains. Emulators with this level of accuracy over such a large volume in parameter space would also facilitate the interchange of results among different collaborations. For example, the DES collaboration primarily uses the \textsc{CosmoSiS} framework~\citep{2015A&C....12...45Z}, while the Rubin observatory Dark Energy Science Collaboration (LSST-DESC) calculates its observable using the \textsc{Core Cosmology Library} and \textsc{FireCrown}~\citep{LSSTDarkEnergyScience:2018yem,LSST:2022sql}. These frameworks are not easily made compatible, and combining data sets at the likelihood level with these codes is cumbersome.

We did not perform exhaustive tests to optimize the size of the training set, aiming for the least amount of points that would still respect minimum quantitative requirements involving $\langle \Delta \chi^2 \rangle$. Training all designs with the masked data vector could also reduce the required training size. We also devoted similar effort to optimizing the hyperparameters in each architecture and adjusting the number and size of each layer to avoid biases when reporting our comparison. We do not claim to have the most optimal hyperparameter solutions, and additional tuning may lower the reported $\Delta \chi^2$ even more. The emulator designs can be combined, and this is precisely what we have done in an accompanying work where we extend the emulator volume of applicability even further.

Increasing the transformer model's complexity can reduce the size of the required training set; it can also preserve the emulator's accuracy when training it on dark energy models with additional free parameters. For instance, we tested an enhanced version of the TFRB emulator where the embedding layer was expanded from twenty-five vectors of dimension 200 to three hundred vectors of dimension 250. In this scenario, we found that the 2x2pt training set could be slightly reduced from two and a half to two million points while maintaining a similar fraction of points with $\Delta \chi^2 < 1$. However, the new emulator is slower to train. These settings should be adjusted according to the available computational resources.

Fig.~\ref{fig:posterior_validation} compares the posterior distributions of key cosmological parameters computed using \textsc{CoCoA} with those obtained from the TFRB, \textsc{ResMLP} and MLP emulators. On the left panel, we assumed the fiducial cosmology to be the $\Lambda$CDM model with parameters located at the center of the prior volume. We adopt the Euclid Emulator version 2.0 to compute the nonlinear matter power spectrum boost in both the $\Lambda$CDM and the growth-geometry split models~\citep{Euclid:2020rfv}. In this split cosmology, we use the growth parameters in the Euclid Emulator, following~\citet{Zhong:2023how}. In this case, Fig.~\ref{fig:posterior_validation} shows, not surprisingly, that the excellent $\Delta \chi^2$ agreement between the TRFB emulator and \textsc{CoCoA} translates into a good agreement at the posterior level. Fig.~\ref{fig:posterior_validation} also shows that the discrepancies at the level $\Delta \chi^2 \sim \mathcal{O}(\mathrm{few})$ seen on the \textsc{ResMLP} emulator do not translate into $\Lambda$CDM parameter biases at the $68\%$ confidence level. There are, however, noticeable differences in the $\log(A_s)$ posterior. As expected, the MLP emulator fails to reproduce accurate results.

\begin{figure*}
\centering
\includegraphics[width=\columnwidth]{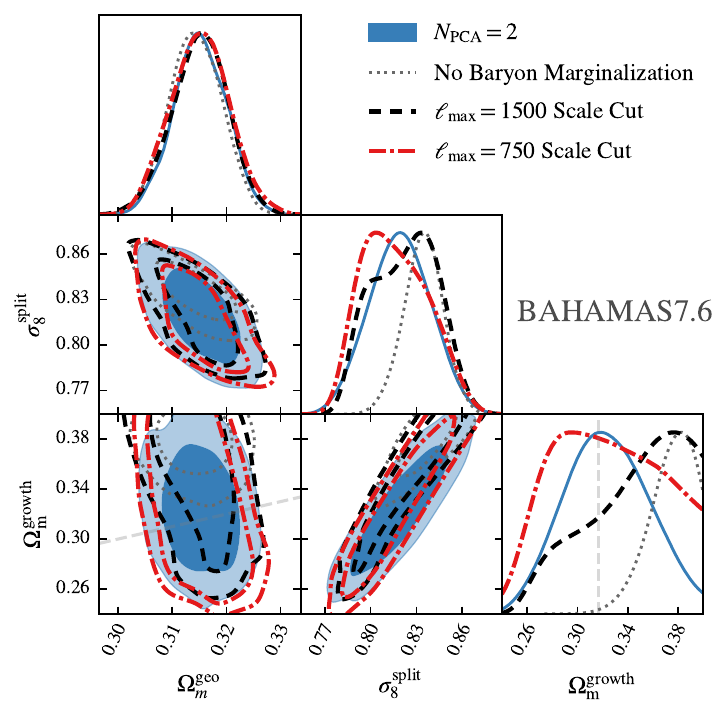}
\includegraphics[width=\columnwidth]{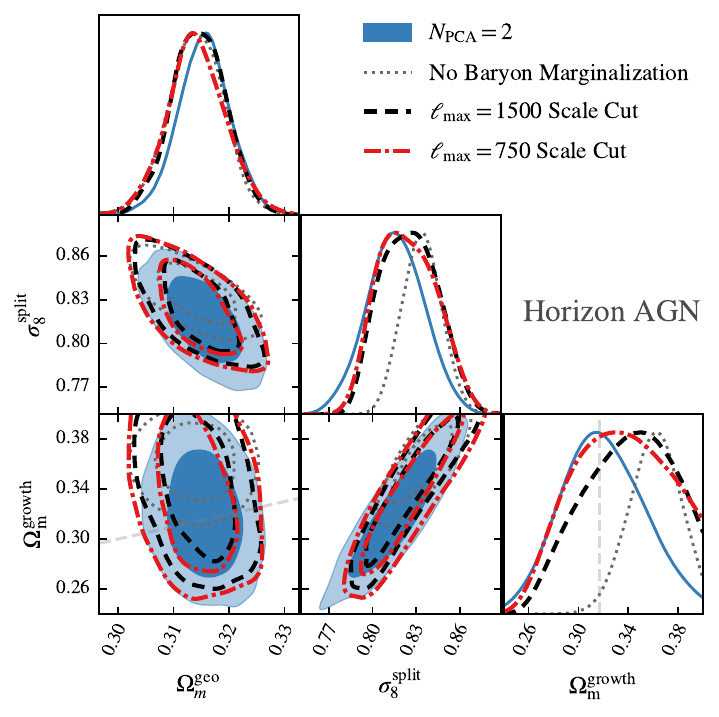}
\caption{Posterior distributions of growth and geometry quantities in the split $\Lambda$CDM model when we introduce baryon feedback contamination on the fiducial cosmic shear data vector. The left panel assumed the feedback derived from the \textsc{BAHAMAS7.6} simulation, while the right panel employed the \textsc{Horizon AGN} scenario. In both panels, the gray short-dashed posteriors are the results of MCMCs in which we have not implemented any technique to mitigate baryonic effects in the theoretical computation of the data vector. On the other hand, the blue solid unbiased posteriors result from chains where we have marginalized the theoretical predictions using two principal components (see equation~\ref{pca_baryon}). The black-dashed and red-dot-dashed posteriors display the results from chains that assumed the more conservative scale cuts $l_\mathrm{max} = 1500$ ($\theta_\mathrm{min} = \{5.51, 17.39\}$ arcmin), and  $l_\mathrm{max} = 750$ ($\theta_\mathrm{min} = \{11.02, 34.78\}$ arcmin) respectively. These angular truncations are based on the first zero-crossing of the Bessel functions that enter into $\xi_{+}$ and $\xi_{-}$ calculations. In all chains, we combined the LSST-Y1 cosmic shear likelihood with the \textsc{All} external data set, which represents Planck 2018 \textsc{clik} low-$\ell$ EE combined with high-$\ell$ TTTEEE ($35<\ell<396$), Type Ia Supernova, BAO and BBN.}
\label{fig:baryon} 
\end{figure*}

The right panel in Fig.~\ref{fig:posterior_validation} displays the results of MCMCs in which the fiducial dark energy possesses anisotropic stresses, following the procedure detailed in Appendix~\ref{sec:ppf}. We expanded the scale cuts, rendering information from nonlinear scales irrelevant. To find such a mask, we first set a fiducial $\Lambda$CDM cosmology and then computed the cosmic shear data vectors with ($\boldsymbol{y}_\mathrm{NL}$) and without ($\boldsymbol{y}_\mathrm{L}$) nonlinear corrections to the matter power spectrum. Then we kept eliminating scales until $\Delta \chi^2_{\mathrm{L-NL}} \equiv \left( \boldsymbol{y}_{\mathrm{NL}} - \boldsymbol{y}_{\mathrm{L}} \right) \textbf{C}^{-1} \left( \boldsymbol{y}_{\mathrm{NL}} - \boldsymbol{y}_{\mathrm{L}} \right)^{T} < 1$ at the fiducial $\Lambda$CDM model. This procedure results in the elimination of all  $\xi_-$ points and the lower limit $\theta_\mathrm{min} = 30$ arcmin for $\xi_+$. After the test on the fiducial model, we run an MCMC chain without any nonlinear corrections to matter power spectrum corrections. We then post-process the parameters that were accepted by the MCMC, turning on the nonlinear gravitational collapse and recomputing the LSST-Y1 likelihood. Finally, we compute the $\Delta \chi^2_{\mathrm{L-NL}}$ distribution to visually inspect whether the scale cuts were adequate to most accepted cosmologies in the MCMC chain.

The exotic fiducial dark energy cosmology with anisotropic stresses shifts the posterior distribution of many cosmological parameters to the edge of the prior. One of the primary motivations for the growth-geometry split is to test whether the substantial disagreement between the geometric dark matter density, $\Omega_\mathrm{m}^\mathrm{geo}$, and the growth dark matter density, $\Omega_\mathrm{m}^\mathrm{growth}$, indicates that the underlying dark energy lies beyond the smooth dark energy paradigm~\citep{Mortonson:2008qy}. The emulators must, therefore, produce sensible results when the fiducial cosmology is far from $\Lambda$CDM. In this case, we see noticeable but not significant discrepancies between the TRFB and \textsc{ResMLP} emulators, with the former always being closer to \textsc{CoCoA} predictions.

\subsection{Bayesian Evidence}\label{sec:bayesian_evidence}

\begin{figure*}
\centering
\includegraphics[width=\columnwidth]{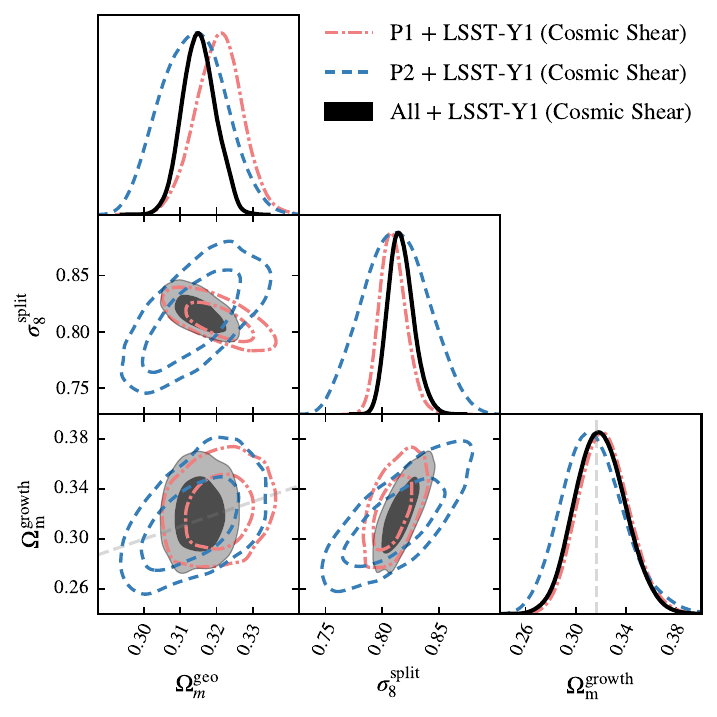}
\includegraphics[width=\columnwidth]{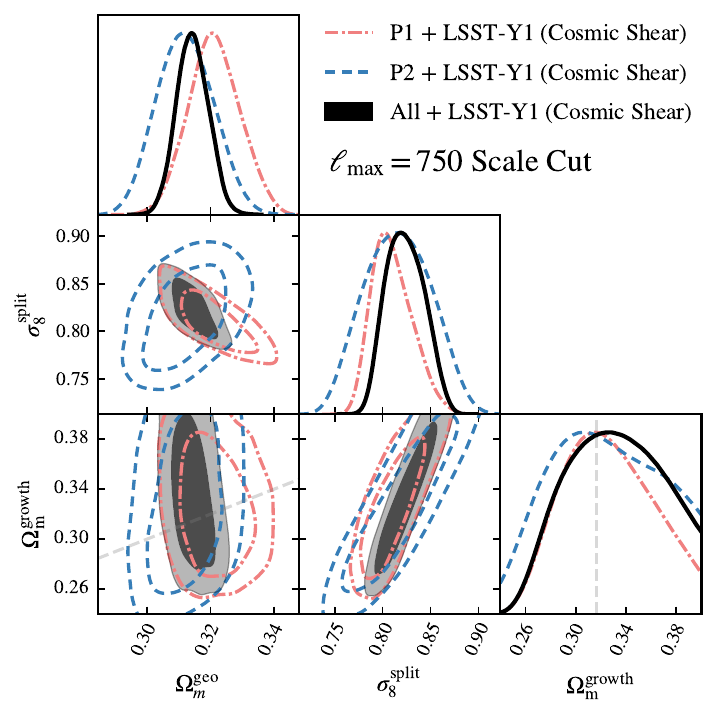}
\caption{Posterior distributions of growth and geometry quantities in the split $\Lambda$CDM model, comparing the effects of external data sets when combined with LSST-Y1 cosmic shear. The chains in the right panel assumed the more conservative $l_\mathrm{max} = 750$ ($\theta_\mathrm{min} = \{11.02, 34.78\}$ arcmin) scale cut. These angular truncations are based on the first zero of Bessel functions that enter into $\xi_{+}$ and $\xi_{-}$ calculations. The P1 external data set represents Planck 2018 \textsc{clik} low-$\ell$ EE combined with high-$\ell$ TTTEEE ($35<\ell<396$). The P2 data set combines Type Ia Supernova, BAO, and BBN (no information on the amplitude and tilt of scalar initial power spectrum). The \textsc{ALL} includes both P1 and P2. In all chains, we did not contaminate the fiducial data vector with baryonic feedback; we also assumed noiseless data vectors. The LSST-Y1 cosmic shear + \textsc{All} results, when compared against  LSST-Y1 cosmic shear + P1, indicates that P2 helps testing the growth-geometry consistency by further constraining $\Omega_\mathrm{m}^\mathrm{geo}$.} 
\label{fig:lcdm_1}
\end{figure*}

The Bayesian evidence is the integral of the product of the likelihood $\mathcal{L}$ and prior $\Pi$ over the parameter space $\theta$:
\begin{equation}
\mathcal{Z}=\int \mathcal{L}(\theta) \times \Pi(\theta) \,  \mathrm{d} \theta.
\end{equation}
The Bayesian evidence is computationally expensive on high dimensional spaces. Evidence is ignored in Monte Carlo Markov chains, being solely a normalizing factor. However, this is a powerful tool for comparing theoretical models and the tension between different data sets~\citep{2006PhRvD..73f7302M,Trotta:2007hy,Trotta:2008qt}.

Nested Sampling~\citep{2004AIPC..735..395S} has been widely used to compute the Bayesian evidence when testing dark energy models with weak lensing and galaxy clustering data~\citep{DES:2017myr,Heymans:2020gsg,DES:2021wwk}. This technique generates samples within nested contours of increasing likelihood, and there are several different algorithms readily available to the community including \textsc{MultiNest}~\citep{2008MNRAS.384..449F}, \textsc{PolyChord}~\citep{2015MNRAS.450L..61H}, \textsc{dynesty}~\citep{2020MNRAS.493.3132S}, \textsc{pocoMC}~\citep{karamanis2022accelerating}, and \textsc{Nautilus}~\citep{Lange:2023ydq}.

\textsc{MultiNest} and \textsc{PolyChord} have been extensively tested in the context of the Dark Energy Survey year one data~\citep{2023MNRAS.521.1184L, Miranda:2020lpk}. According to these studies, \textsc{PolyChord} produces stable results that are less sensitive to changes in the multiple hyperparameters that control their accuracy. Therefore, we adopt \textsc{PolyChord} in this work with the following conservative hyperparameters settings: $n_\mathrm{live} = 1024$, \textsc{precision criterion} $= 0.001$, and $n_\mathrm{repeat} = 5 d$, where $d$ is the dimensionality of the parameter space.

The number of sampling points necessary for convergence varies between nested sampling methods, but they all seem to demand computational resources larger than what is typically required for MCMC convergence. While we are not altering the relative cost between nested samplers and Metropolis-Hasting algorithms, emulators reduce the absolute CPU expense of all statistical methods. However, Bayesian evidence computation with emulators requires them to have a large volume of reasonable accuracy. Poor or non-existent training can induce emulators to have nonsensical results due to the low-likelihood regions neglected in MCMCs (for instance, a long, flat three-sigma tail in the posterior distribution). These regions can still affect the final evidence value if they are associated with enough prior volume.

We validate the accuracy of our emulators in computing the Bayesian evidence in three distinct cosmologies, one being at the center of the cosmology prior box. We summarize the results in Table~\ref{table:bayesian_evidence}. The TRFB is the only emulator accurate within the error bars of the \textsc{PolyChord} precision. Once more, the baseline emulator trained with 2M points performs generally better than \textsc{ResMLP}. For \textsc{ResMLP}, inaccuracies become more noticeable when the posterior distribution of cosmological parameters is closer to the prior boundaries. While adding an extra few million points could, in principle, make \textsc{ResMLP} perform similarly to the baseline TRFB emulator on cosmic shear, we expect the comparison to be more challenging when dealing with 3x2pt data vectors. However, we leave such detailed testing for future work.

Both the computational expenditure and the execution time required to compute Bayesian evidence were a lot smaller using the TRFB emulator compared to \textsc{CoCoA}, by more than one order of magnitude. Such a significant difference creates new scientific opportunities, including the comparison of samplers and the reproduction of past work while spending minimal computational resources. For instance,~\citet{2023MNRAS.521.1184L} run twenty different precision settings for each $N_{\rm live}$ choice, while~\citet{Miranda:2020lpk} runs \textsc{Polychord} hundreds of times to calibrate the expectation value of Bayesian evidence ratios given the DES year one prior choices. 

\section{The Growth-Geometry split}\label{sec:gg-split}

The split matter power spectrum and the corresponding $\sigma_b^\mathrm{split}$ are defined as
\begin{align}\label{def:Pk-slpit}
\begin{aligned}
P_{\rm split}(k , z) &=  P_{\rm geo}^{\rm linear}(k,z) \times  \bigg(\dfrac{G_{\rm growth}^{\rm ODE}(z)}{G_{\rm geo}^{\rm ODE}(z)}\bigg)^2  \times B^{\rm growth}(k, z), \\
\bigg(\sigma_{8}^{\mathrm{split}}(z) \bigg)^2 &=\frac{1}{2 \pi^{2}} \int \mathrm{d} \log k W_8^{2}(k R) k^{3} P^{\mathrm{split}}(k, z),
\end{aligned}
\end{align}
where $B^{\rm growth}$ is the boost factor from the \textsc{Euclid Emulator} v2.0~\citep{Euclid:2020rfv} that accounts for the non-linear contribution based on growth parameters. The linear power spectrum is rescaled according to the growth function $G(z)$ that obeys
\begin{align}\label{eq:g-growth}
G^{\prime \prime}+ \left(4+\frac{H^{\prime}}{H}\right) G^{\prime}+\left[3+\frac{H^{\prime}}{H}-\frac{3}{2} \Omega_{\mathrm{m}}(z)\left(1-f_\nu\right)\right] G=0. 
\end{align}
Here, prime denotes the derivative with respect to the logarithm of the scale factor $\log a$, and $f_\nu = \Omega_\nu/\Omega_\mathrm{m}$ is the fraction of the neutrino mass. The correction for neutrino mass follows the fitting formula in~\citet{Kiakotou:2007pz} to account for the relative change in the growth function, which is not included in~\citet{Zhong:2023how}. When the neutrino mass is fixed to $\sum_{\nu} m_{\nu} = 0.06\mskip\thinmuskip \mathrm{eV}$, the impact of this correction is small as $f_\nu \ll 1$. In this paper, we assume the minimum neutrino mass scenario. 

\subsection{Forecasting Results}

We adopt the TRFB emulator in all our forecasts. We combine the LSST-Y1 cosmic shear and 3x2pt simulated data with three choices of external experiments that do not provide additional information on the growth parameters. The first choice, denoted \textsc{P1}, is a restricted version of the primary Cosmic Microwave Background Temperature and Polarization data~\citep{Zhong:2023how,Xu:2023qmp}. \textsc{P1} includes the Planck 2018 \textsc{clik} low-$\ell$ EE polarization likelihood combined with Planck 2018 \textsc{clik} high-$\ell$ TTTEEE on the restricted $35<\ell<396$ range~\citep{Planck:2019nip}. The selected multipoles prevent the gravitational lensing of the CMB from influencing the parameter posteriors. Similarly, the absence of the Planck 2018 low-$\ell$ TT data impedes the late-time integrated Sachs-Wolf effect from constraining the growth parameters~\citep{Hu:1994jd,Hu:1995kot}. 

The second geometry prior choice, denoted \textsc{P2}, comprises the combination of Type Ia Supernova data from the Pantheon sample~\citep{Pan-STARRS1:2017jku} with baryon acoustic oscillations (BAO) and Big Bang Nucleosynthesis (BBN) constraints. The BAO data combines the SDSS DR7 main galaxy sample \citep{Ross:2014qpa}, the 6dF galaxy survey at $z_{\text{eff}} = \{0.106,0.15\}$~\citep{beutler20116df}, and the SDSS BOSS DR12 low-z and CMASS galaxy samples at $z = \{0.38, 0.51, 0.61\}$ \citep{BOSS:2016wmc}. \textsc{P2} also includes the prior $100 \Omega_\mathrm{b}h^2 = 2.208\pm0.052$ derived from BBN~\cite{DES:2017txv}. Unlike P1, the P2 set provides no external information on the amplitude $\log(A_s)$ and tilt  $n_s$ of the scalar initial power spectrum. Finally, we introduce the \textsc{All} combination as a third option, which includes both P1 and P2.

\begin{figure}
\centering
\includegraphics[width=\columnwidth]{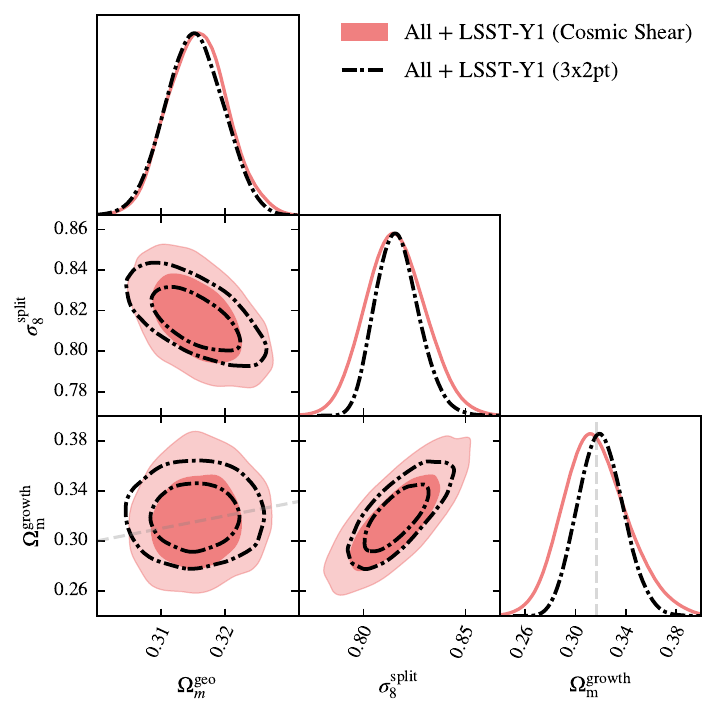}
\caption{Posterior distributions of growth and geometry quantities in the split $\Lambda$CDM model, comparing constraints from LSST-Y1 cosmic shear and LSST-Y1 3x2pt simulated likelihoods, both combined with the \textsc{All} external data set and marginalized over baryonic effect. We found an approximately $30\%$ improvement in the constraining of the growth parameter $\Omega_\mathrm{m}^{\mathrm{growth}}$ by including galaxy-galaxy lensing and galaxy clustering. In both cases, we assumed noiseless data vectors.} 
\label{fig:lcdm_3x2}
\end{figure}

The adopted fiducial scale cuts are sensitive to baryonic physics, as seen in Fig.~\ref{fig:baryon}. Fortunately, there are several approaches to attenuate baryonic effects on cosmic shear~\citep{2015MNRAS.450.1212H, 2021MNRAS.502.1401M, 2019JCAP...03..020S, 2020MNRAS.495.4800A}. Our study uses a well-established mitigation scheme based on principal components constructed from the effects that different scenarios of baryonic feedback imprint on the matter power spectrum at a few cosmologies; the assumption here is that these effects are not strongly correlated with cosmology~\citep{hem19, 2021MNRAS.502.6010H, 2014MNRAS.440.1379E}. Recent studies~\citep{2021MNRAS.502.6010H} have shown that the marginalization with respect to the leading two eigenvectors, constructed from eight different feedback scenarios, is enough to unbias $\Lambda$CDM posteriors. Accordingly, we only added the two leading principal components, $\{\vec{e}_1, \vec{e}_2\}$, to our MCMC chains. 

The principal components change the cosmic shear data vector with a linear addition that does not require our emulator to be retrained
\begin{equation}\label{pca_baryon}
\xi_{ \pm} \rightarrow \xi_{ \pm} + \sum_{i=1}^2 Q_i \vec{e}_i.
\end{equation}
We utilized eight hydrodynamical simulations to construct the principal components (eigenvectors): \textsc{Illustris TNG-100}~\citep{2018MNRAS.475..676S}, \textsc{MassiveBlack-II}~\citep{2015MNRAS.450.1349K}, \textsc{Horizon-AGN}~\citep{2014MNRAS.444.1453D}, two simulations (\textsc{T8.0}, \textsc{T8.5}) from \textsc{Cosmo-OWLS} suite~\citep{2014MNRAS.441.1270L}, and three simulations (\textsc{T7.6}, \textsc{T7.8}, \textsc{T8.0}) from the \textsc{BAHAMAS} suite~\citep{2017MNRAS.465.2936M}. For the coefficients $Q_i$, we adopted the conservative prior $\Pi(Q_i) = \textsc{Flat}[-100,100]$.

In Fig.~\ref{fig:baryon}, we test the efficacy of the principal component method, restricted to two additional components, in unbiasing the cosmological parameters. We focus on the geometry and growth cold dark matter density as well as the $\sigma_8^\mathrm{split}$ amplitude. In the left (right) panel, we contaminated the fiducial data vector with baryonic feedback from \textsc{BAHAMAS T7.6} (\textsc{Horizon AGN}) simulations. Fig.~\ref{fig:baryon} also compares the PCA marginalization against chains with fixed $Q_{i=1,2} = 0$, and imposed $l_\mathrm{max} = 700, 1500$ cuts. The angular cuts corresponding to $l_\mathrm{max} = 1500$ are $\theta_\mathrm{min} = \{5.51, 17.39\}$ arcmin for $\left\{\xi_{+}, \xi_{-}\right\}$, while for $l_\mathrm{max} = 700$, they are  $\theta_\mathrm{min} = \{11.02, 34.78\}$ arcmin (these truncations are based on the first zero of Bessel functions that enter on $\xi_{+}$ and $\xi_{-}$).

The posteriors in Fig.~\ref{fig:baryon} suggest the PCA method is superior, compared to applying conservative scale-cuts, in correcting biases in the $\Omega_\mathrm{m}^\mathrm{growth}$ parameter without excessive loss of constraining power. We did not explore scales within the range $700 < l_\mathrm{max} < 1500$, or further test the PCA method against simulations not used in constructing them. However, Fig.~\ref{fig:baryon} reveals a sharp transition from significantly biased results with $l_\mathrm{max} = 1500$ to overly-broad posterior when $l_\mathrm{max} = 700$. Therefore, determining a unique $l_\mathrm{max}$ that optimizes constraints to be tighter than what PCA can accomplish while being robust against all possible scenarios may not be possible.

Fig.~\ref{fig:lcdm_1} presents $\Lambda$CDM-split forecasts, combining LSST-Y1 cosmic shear with \textsc{P1}, \textsc{P2}, or \textsc{All} external data sets. In the left panel, we maintain the fiducial scale cuts and marginalize over the parameters $Q_{i=1,2}$. The right panel employs the conservative $l_\mathrm{max} = 700$ cut, with fix $Q_{i=1,2} = 0$. In all chains, we did not contaminate the fiducial data vector with baryonic feedback; we also assumed noiseless data vectors. Albeit these are consequential simplifications, Fig.~\ref{fig:lcdm_1} does show the importance of CMB data in constraining early universe parameters that do correlate with $\Omega_\mathrm{m}^\mathrm{growth}$.  The LSST-Y1 + \textsc{All} results, when compared against  LSST-Y1 + P1, indicate that P2 helps testing the growth-geometry consistency by further constraining $\Omega_\mathrm{m}^\mathrm{geo}$.

The comparison between the left and right panels in Fig.~\ref{fig:lcdm_1} reveals the large impact that small scales have in constraining $\Omega_\mathrm{m}^\mathrm{growth}$. With the fiducial cosmic shear scale cut, the uncertainty on $\Omega_\mathrm{m}^{\mathrm{growth}}$ is reduced by about $53\%$ on LSST-Y1 cosmic shear + \textsc{All} compared with the DES-Y3 cosmic shear + \textsc{All}~\citep{Zhong:2023how}. Compared to DES-Y3 3x2pt + \textsc{All}, the uncertainty on $\Omega_\mathrm{m}^{\mathrm{growth}}$ is reduced by about $42\%$. In contrast, the conservative scale cut weakens the constraining power on LSST-Y1 cosmic shear + \textsc{All} forecast to the level of DES-Y3 3x2pt + \textsc{All}. The caveat is that~\citet{Zhong:2023how} did not explore such an aggressive scale cut for the $\xi_-$ component. Nevertheless, our analysis demonstrates the potential improvements that are theoretically achievable with LSST-Y1 data.

Fig.~\ref{fig:lcdm_3x2} compares constraints on $\Lambda$CDM-split parameters derived from LSST-Y1 cosmic shear and 3x2pt, each incorporating the \textsc{All} external data set. The galaxy-galaxy lensing had simplifying assumptions; for instance, we assumed linear galaxy biases instead of higher-order and more precise prescriptions such as Hybrid Effective field theories~\citep{Kokron:2021xgh,Nicola:2023hsd}. This choice prevented us from investigating the effect of more aggressive $R_\mathrm{min}$ cuts on the split posteriors. However, this approach significantly simplified the training process for our 2x2pt machine learning emulator, as linear biases are fast parameters that only renormalize the data vector. We did not include baryonic feedback or noise in the fiducial data vector but we sampled the 2 PCs to marginalize the baryonic effect. Nonetheless, the \textsc{All} + LSST-Y1 3x2pt chain showed $\Omega_\mathrm{m}^{\mathrm{growth}}$ error bars that were  $30\%$ smaller compared to \textsc{All} + LSST-Y1 cosmic shear.

\subsection{Limits of the Growth-Geometry Split}\label{sec:limit_of_ggsplit}

\begin{figure}
\centering
\includegraphics[width=\columnwidth]{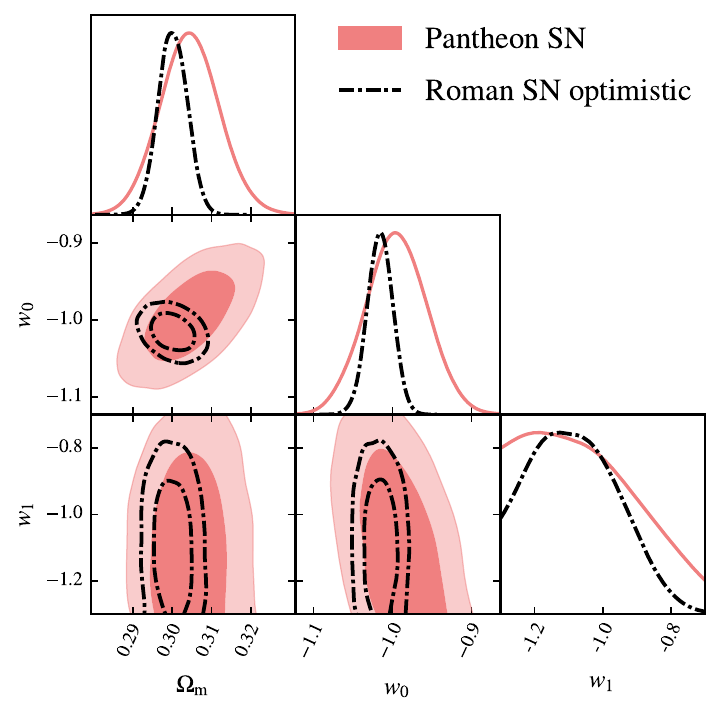}
\caption{Posterior distributions for the late-time smooth dark energy $(w_0, w_1)$ parameters in the two-bin model defined on equation~\ref{eq:wz_params}. The red solid lines include Pantheon Type Ia supernova real data, while the black dot-dashed line replaces Pantheon with Roman space telescope simulated data (the optimistic \textsc{Imaging-Allz} synthetic sample from~\citet{2018ApJ...867...23H}). Both cases also include the 2018 Planck temperature and polarization data (\textsc{clik} likelihood, full multipole range) and BAO (SDSS DR7 main galaxy sample, the 6dF galaxy survey, and the SDSS BOSS DR12 low-z and CMASS galaxy samples). We utilized these posteriors to produce the growth-geometry mapping discussed in Sec.~\ref{sec:limit_of_ggsplit}.}
\label{fig:wbin_posterior} 
\end{figure}

The main goal of the growth-geometry split is to provide a straightforward phenomenological test, without an excessive number of parameters, of the validity of the smooth paradigm for cosmic acceleration~\citep{Mortonson:2008qy}. Late-time dark energy models consistent with this set of theories affect the linear growth of structures only by changing the expansion rate in eq.~\ref{eq:g-growth}. Detection of additional scale dependency or an anomalous redshift evolution in the matter power spectrum would open new theoretical possibilities. In the growth-geometry split, the expectation is that large deviations from the smooth dark energy would trigger the split parameters to be strongly inconsistent, exemplified in the right panel of Fig.~\ref{fig:posterior_validation} as the presence of anisotropic stresses is a possible alternative to explain existing tensions outside the smooth paradigm~\citep{Koivisto:2005mm,Cardona:2014iba,Majerotto:2015bra,2020PDU....3000668G}.

The $\Lambda$CDM split is the simplest incarnation of the growth-geometry split, as it only duplicates the cold dark matter density. Having a single additional parameter provides sensitivity to the consistency test, as both $\Omega_\mathrm{m}^{\rm geo}$ and $\Omega_\mathrm{m}^{\rm growth} $ parameters can be tightly constrained. The drawback of this simplification lies in the fact that a more general late-time dark energy equation of state can trigger $\Omega_\mathrm{m}^{\rm geo} \neq \Omega_\mathrm{m}^{\rm growth}$. 

A possible strategy for testing whether detected deviations in the split parameters indicate new physics beyond smooth dark energy involves an iterative procedure. If the posterior of $\Lambda$CDM split parameters exhibit inconsistencies, then increase the number of parameters by dividing $w(z) = \mathrm{const}$ into  $w^\mathrm{geo}$ and $w^\mathrm{growth}$ components. Subsequently, if the wCDM-split consistency also fails, then increase the number of parameters again by introducing time dependency in the geometry and growth components of the dark energy equation of state. Such an iterative approach must be accompanied by Bayesian model comparison to check whether the additional parameters are justified in fitting the geometry data; otherwise, the degrading constraining power may erase the growth-geometry split usefulness~\citep{Trotta:2008qt,Liddle:2009xe}.

\begin{figure*}
\centering
\includegraphics[width=2\columnwidth]{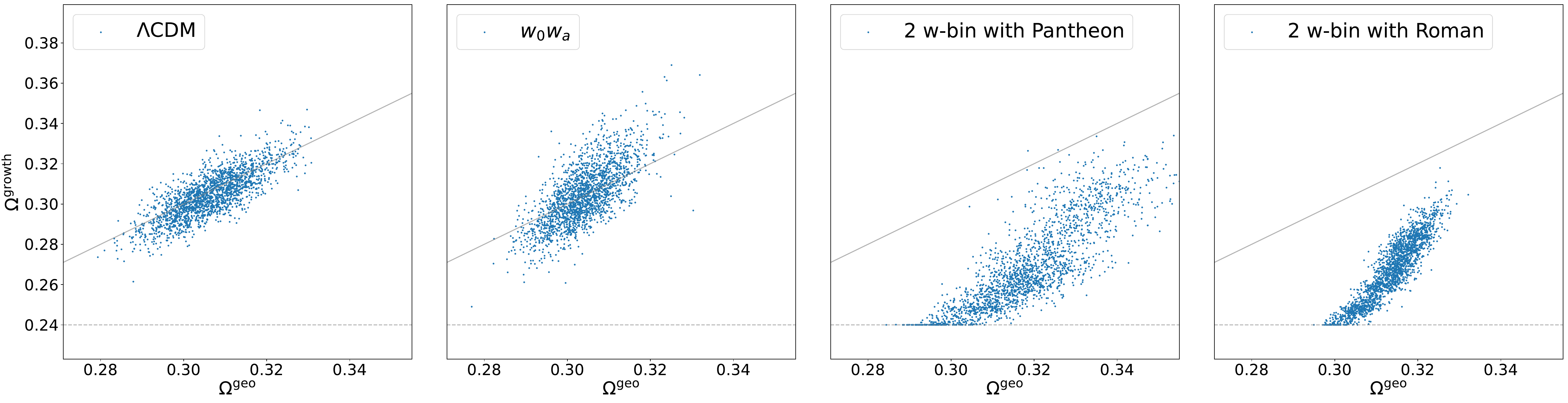}
\caption{Best fit $\Omega_\mathrm{m}^{\rm geo}$ and $\Omega_\mathrm{m}^{\rm growth}$ parameters in the $\Lambda$CDM split when the underlying late-time dark energy model is a more general smooth dark energy model. We ran MCMC chains that assumed three illustrative $w(z)$ parameterizations: $\Lambda$ (first panel), CPL (second panel) or two-parameter piecewise function (third panel). These runs constrained the cosmological parameters by combining the baseline high-$\ell$ TTTEEE, low-$\ell$ Planck power spectra, Type Ia supernova, and BAO data, following~\citet{Planck:2018vyg}.  We also considered one additional chain for the piecewise $w(z)$, in which we swap the Pantheon supernova data with the optimistic \textsc{Imaging-Allz} synthetic sample from~\citet{2018ApJ...867...23H} (fourth panel). We computed the LSST-Y1 cosmic shear data vector for every accepted parameter in the MCMC chains using \textsc{Cosmolike}. For simplicity, we fixed nuisance parameters to the fiducial values displayed in Tab.~\ref{table:priorchoices}. Finally, we minimized the LSST-Y1 likelihood to find the best $\Lambda$CDM-split parameters that approximate each model. When computing the non-linear matter power spectrum in the piecewise model, we adopted the Casarini prescription that maps the boost factor of any $w(z)$ to a $w$CDM model. The $\Lambda$ investigation acted as a null test of our pipeline. The observed clustering of points around the $\Omega^{\rm geo}_\mathrm{m} = \Omega^{\rm growth}_\mathrm{m}$ line in the second panel is positive news as CPL encompasses what most scalar field models should predict. 
}
\label{fig:mapping_1} 
\end{figure*}

Another approach would use the most comprehensive available data to find the best, again taking into account Bayesian model comparison techniques, $w(z)$ parameterization from theoretically motivated or data-driven functional forms. Then, we split the parameters that constitute the best-fit $w(z)$, hoping the available growth data can provide sufficient constraining power for the multiple growth-related free parameters. Unfortunately,~\citet{DES:2020iqt} and \citet{Zhong:2023how} have indicated that there is no sufficient information in the DES 3x2pt year one and three data to constrain all growth parameters even in the wCDM split.

We propose a third solution that significantly profits from the speed enhancements our TRFB emulator offers. We advocate growth-geometry tests to be limited to the $\Lambda$CDM-split. Then, viable smooth dark energy models with more general $w(z)$ should be mapped into the $\big(\Omega_\mathrm{m}^{\rm geo}, \Omega_\mathrm{m}^{\rm growth} \big)$ plane in the $\Lambda$CDM-split. As an example, we considered two time-dependent $w(z)$ parameterizations. The first model was the Chevallier, Polarski, and Linder (CPL)  $w(z) = w_0 + w_a z/(1+z)$~\citep{Chevallier:2000qy,Linder:2002et}. The second parameterization was the piecewise equation of state defined as follows:
\begin{equation}\label{eq:wz_params}
    w(z) = \begin{cases}
        w_0, & \text{ if } 0 \leq z < 1.0\\
        w_1, & \text{ if } 1.0 \leq z < 2.0\\
       -1, & \text{ if } z \geq 2.0.
    \end{cases}
\end{equation}
The piecewise $w(z)$ provides much greater freedom in the $1.0 \leq z < 2.0$ range that current Type Ia supernova and BAO data have not been well covered; therefore, it serves well as an extreme test of the growth-geometry split.

We ran MCMC chains that assumed piecewise late-time smooth dark energy. These runs constrained the cosmological parameters by combining the baseline high-$\ell$ TTTEEE, low-$\ell$ Planck power spectra, Type Ia supernova, and BAO data, following~\citet{Planck:2018vyg}. In the CPL model, we use the chains retrieved directly from the Planck Collaboration public archive\footnote{\hyperlink{https://www.cosmos.esa.int/web/planck}{https://www.cosmos.esa.int/web/planck}}~\citep{Planck:2018vyg}. We also considered one additional chain for the piecewise $w(z)$, in which we swap the Pantheon supernova data with the optimistic \textsc{Imaging-Allz} SN synthetic sample from~\citet{2018ApJ...867...23H} that simulates the capabilities of the Roman Space Telescope. Fig.~\ref{fig:wbin_posterior} compares their constraining power on $w_0$ and $w_1$ parameters, showing that Roman will be able to better probe the high redshift behavior of $w(z)$ (the rise of dark energy as explained in~\citet{Linder:2021syd,Linder:2023klx}). The \textsc{Imaging-Allz} models the SN samples that the Roman Space Telescope could observe in the redshift range $0 < z < 3$ with only photometry, assuming the most favored scenarios on the mitigation of systematic effects. 

We compute the LSST-Y1 cosmic shear data vector for every accepted parameter in the MCMC chains using \textsc{Cosmolike}. For simplicity, we fixed nuisance parameters to the fiducial values displayed in Table.~\ref{table:priorchoices}. We can generalize this approximation in the future by assigning nuisance parameters that minimize the LSST-Y1 likelihood, which will require constructing emulators for these beyond wCDM cosmologies. When computing the non-linear matter power spectrum in the piecewise model, we adopted the Casarini prescription that maps the boost factor of any $w(z)$ to a $w$CDM model~\citep{2009JCAP...03..014C}. Finally, we minimize the LSST-Y1 likelihood to find the best $\Lambda$CDM-split parameters that approximate each model.

For the CPL $w(z)$,  the second panel from the left in Fig.~\ref{fig:mapping_1} reveals that the growth-geometry $\Lambda$CDM split does not induce an extreme   $\Omega^{\rm geo}_\mathrm{m} \neq \Omega^{\rm growth}_\mathrm{m}$ incompatibility. This clustering of points around the $\Omega^{\rm geo}_\mathrm{m} = \Omega^{\rm growth}_\mathrm{m}$ line is good news because most physical scalar field models can be reasonably mapped to the CPL parameterization. However, splitting only the matter density is no longer sufficient when the background evolution is further relaxed to the more radical two-bin $w(z)$ parameterization, as shown in the third panel of Fig.~\ref{fig:mapping_1}. The last panel in Fig.~\ref{fig:mapping_1} shows that the additional constraining power from the Roman Space Telescope clusters the points even more. This result demonstrates the usefulness of machine learning emulators with a large volume of parameter space.

\section{Conclusion}\label{sec:conclusion}

We built a machine-learning emulator that accurately produces, over a large volume in parameter space, the real space cosmic shear, galaxy-galaxy lensing, and galaxy clustering correlation functions for the growth-geometry split in the context of LSST-Y1. Emulators provide a realistic approach to addressing concerns on the growing computational requirements for employing current and future large-scale structure surveys in falsifying cosmological models; they can also help in the comprehensive synthetic analyses that collaborations must perform to weigh the modeling options before any data is unblinded~\citep{DES:2017tss,DES:2021rex}. 

The TRFB emulator can be seen, in some aspects, as an upgrade to the proposed machine learning-based accelerator developed in~\citet{Boruah:2022uac}. Our emulator has a range of applicability in parameter volume analogous to the one adopted by the \textsc{Euclid Emulator v2.0}. Our training method sample points using a mixture of LHS and uniform samplers, requiring a few million points for the emulator to achieve precision of the order of $\Delta \chi^2 \sim 0.2$. This number is significantly higher than the training requirement in~\citet{Boruah:2022uac}, revealing that their iterative approach is still suitable. The trivial parallelization of our sampling method can reduce the required training execution time to a few hours, given enough resources. 

Nonetheless, the additional complexity of our design, which contains four times more trainable parameters than the simple MLP, drove us to adopt state-of-the-art GPUs that may not be readily available everywhere, again pointing to the applicability of~\citet{Boruah:2022uac} in simple forecasting applications. The precise volume to which we can accurately train our proposed emulator depends on the number of free parameters in the dark energy model; the same is true for the required number of training points. In this regard, we have not tested how many additional parameters we can include in the cosmological model while maintaining the range of $\Lambda$CDM parameters to be the one given by the Euclid Emulator and still have only a small fraction of validation points with $\Delta \chi^2 > 1$. However, the relative performance of the machine-learning emulators we tested should hold regardless of the number of free parameters in the dark energy model.

We provide a detailed comparison between three machine-learning designs that illustrate the importance of adopting the state-of-the-art Transformer block when constructing emulators. Equal effort was devoted to optimizing the hyperparameters in each architecture and fine-tuning the number and sizes of layers, ensuring an unbiased comparison. We show in Fig.~\ref{fig:chi2_list} that the TRFB emulator improves the overall fit to the \textsc{CoCoA} calculations by one order of magnitude compared to \textsc{ResMLP} and two orders of magnitude compared to \textsc{MLP}. Our tests reveal that increasing the transformer model's complexity can reduce the required training set at the expense of longer training times that consume way more GPU resources.

Our detailed study portrayed how different emulator designs performed when computing Bayesian evidence for cosmic shear on three fiducial cosmologies. The evidences were computed using the \textsc{PolyChord} nested sampler with conservative hyperparameters. We investigate the impact of training set sizes on Bayesian evidence accuracy for each design, considering three scenarios. The TRFB was the only emulator accurate within the error bars of the \textsc{PolyChord} precision for all training sizes and cosmologies. Nevertheless, the \textsc{ResMLP} did not perform poorly when trained with 8 million points. While we have not performed a similar set of comparisons for the 3x2pt emulator, fig.~\ref{fig:chi2_list2} shows an excellent $\Delta \chi^2$ agreement, taking into account the full 3x2pt covariance matrix, between TRFB and the exact calculation. Based on these findings, we anticipate that the TRFB will also be able to compute the Bayesian evidence accurately in this scenario.

As an application, the baseline TRFB emulator was employed to forecast how well the upcoming Rubin Observatory will be able to probe the consistency between  $\Omega^{\rm geo}_\mathrm{m}$ and  $\Omega^{\rm growth}_\mathrm{m}$ in the split $\Lambda$CDM scenario. The chosen fiducial scale cuts were sensitive to baryonic effects, as shown in Fig.~\ref{fig:baryon}. However, we show that principal components constructed from the effects that a few selected scenarios of baryonic feedback imprint on the matter power spectrum at a few fixed cosmologies are promising in unbiasing $\Omega^{\rm growth}_\mathrm{m}$ posteriors. Comparing the PC method against more conservative scale cuts on cosmic shear indicated that the former provides more constraining error bars on  $\Omega^{\rm growth}_\mathrm{m}$. 

Fig.~\ref{fig:lcdm_1} shows forecasts on how cosmic shear can be combined with three distinct combinations of external data that do not provide information on the growth parameters. Incorporating external information on $\log(A_s)$ and  $n_s$ contributed to reducing the error bars on the growth parameter. We also compared cosmic shear against 3x2pt in Fig.~\ref{fig:lcdm_3x2}, with the 3x2pt inference reducing $\Omega^{\rm growth}_\mathrm{m}$ uncertainties by approximately $30\%$.  

Finally, Fig.~\ref{fig:posterior_validation} shows that late-time dark energy models outside the smooth paradigm of cosmic acceleration can create strong inconsistencies between $\Omega^{\rm geo}_\mathrm{m}$ and  $\Omega^{\rm growth}_\mathrm{m}$. This was observed even when the analysis was limited to cosmic shear with scales only sensitive to the linear gravitational evolution of the matter power spectrum. However, general $w(z)$ parameterizations can also yield false positives in split $\Lambda$CDM studies. Fortunately, we were able to show that viable models in the CPL $w(z)$ do not induce large variations in the difference between geometry and growth densities.

There are several avenues in which this work can be improved, including expanding the still limited volume of parameter space that the emulator can be applied with high accuracy. The \textsc{Euclid Emulator} parameter box is informative in statistical inferences that simulate the capabilities of the Rubin observatory. In subsequent work, we will explore alternative designs that integrate \textsc{ResMLP} and TRFB, as well as different training methods that will expand the volume to such an extent that the new priors will be uninformed even in the context of the Dark Energy Survey. Another improvement relates to pushing the 3x2pt toward small scales using better-motivated galaxy biases and intrinsic alignment models from an effective field theory perspective. We leave that for a future investigation. 

\section*{Acknowledgements}
We would like to thank Eduardo Rozo for useful discussions. The authors would also like to thank Stony Brook Research Computing and Cyberinfrastructure, and the Institute for Advanced Computational Science at Stony Brook University for access to the high-performance SeaWulf computing system, which was made possible by the National Science Foundation grant $\#1531492$. TE is supported by the Department of Energy grant DE-SC0020215 and DE-SC0023892. EK is supported by the Department of Energy grant DE-SC0020247 and an Alfred P. Sloan Research Fellowship.

\bibliographystyle{mnras}
\bibliography{ref}

\appendix

\section{Non-smooth dark energy Injection}\label{sec:ppf}

This Appendix describes how we generated LSST-Y1 data vectors with non-smooth dark energy that contained anisotropic stresses. We employed the Parameterized Post-Friedmann Framework (PPF)~\citep{Hu:2007pj, Fang:2008sn, Hu:2008zd} because the fluid description suffers from singularities when the dark energy equation of state crosses $w=-1$. The PPF parameterization replaces the fluid equation, and thus avoid the singularities, by establishing a direct correspondence between dark energy fluid and dark matter in the large-scale limit. It also introduces an effective sound speed $c_\Gamma$ to describe perturbations in the small scale limit. PPF introduces a form for interpolating the large and small scale limit~\citep{Hu:2004kh}:
\begin{equation}
\lim _{k_H \ll 1} \Gamma^{\prime}=S-\Gamma\, .
\end{equation}
Here, $k_H = k/aH$ and the source term is
\begin{equation}
\begin{aligned}
S= & \frac{g^{\prime}-2 g}{g+1} \frac{\Phi - \Psi}{2}-\frac{4 \pi G}{(g+1) k_H^2 H^2}\left\{g\left[\left(p_T \pi_T\right)^{\prime}+p_T \pi_T\right]\right. \\
& \left.-\left[\left(g+f_\zeta+g f_\zeta\right)\left(\rho_T+p_T\right)-\left(\rho_e+p_e\right)\right] k_H V_T\right\} \,  .
\end{aligned}
\end{equation}
In this description, the subscript $T$ stands for other components: matter, radiation, and neutrinos. 

The function $\Gamma$ is related to the rest frame dark energy density perturbation via the Poisson equation:
\begin{equation}
k^2 \Gamma=-4 \pi G a^2 \delta \rho_{\mathrm{DE}}^{\mathrm{rest}} \, .
\end{equation}

The free functions in PPF is reduced to a scale and time-dependent metric fluctuation $g(\ln a, k)$ and a time-only-dependent function $f_\zeta(a)$. Note that they can be mapped to specific modified gravity models such as $f(R)$ or DRP~\citep{Hu:2007pj}. We follow the phenomenological parameterization~\citep{Hu:2008zd}:
\begin{equation}
g(\ln a, k)=g_0\left(\frac{\rho_e}{\rho_T} \frac{\Omega_T}{\Omega_e}\right)^{1 / 2} \left( 1+ (c_g k_H)^2 \right)^{-1} .
\end{equation}
and
\begin{equation}
f_\zeta(a)(\ln a)=0.4 g_0\left(\frac{\rho_e}{\rho_T} \frac{\Omega_T}{\Omega_e}\right)^{1 / 2}.
\end{equation}
Previous studies have found, using the full shape Planck 2015 CMB temperature and polarization data, constraints on $g_0$ centered at zero (no anisotropic stress) with a standard deviation $\sigma_{g_0} \approx 0.2$~\citep{2020PDU....3000668G}. For this study, we chose an extreme case $(g_0, c_g) = (-1, 0.01)$ to exemplify the impact of such models on the $\Lambda$CDM growth-geometry split parameters. Finally, we replace the matter power spectrum with the Weyl potential auto spectrum on cosmic shear~\citep{DES:2022ygi, 2019MNRAS.490.2155S}.

\bsp	
\label{lastpage}
\end{document}